\documentclass[journal]{IEEEtran}
\pdfoutput=1
\usepackage{cite}[unsrt]

\ifCLASSINFOpdf
  \usepackage[pdftex]{graphicx}
  \DeclareGraphicsExtensions{.pdf,.jpeg,.png}
\else
  \usepackage[dvips]{graphicx}
  \DeclareGraphicsExtensions{.eps}
\fi
\ifCLASSOPTIONcompsoc
 \usepackage[caption=false,font=normalsize,labelfont=sf,textfont=sf]{subfig}
\else
 \usepackage[caption=false,font=footnotesize]{subfig}
\fi

\usepackage{xcolor}
\usepackage{physics}
\usepackage{ragged2e}
\usepackage{amsmath}
\usepackage{longtable}
\usepackage{supertabular}
\usepackage{capt-of}
\usepackage{scalerel}
\usepackage{tikz}
\usetikzlibrary{svg.path}
\allowdisplaybreaks

\definecolor{orcidlogocol}{HTML}{A6CE39}
\tikzset{
  orcidlogo/.pic={
    \fill[orcidlogocol] svg{M256,128c0,70.7-57.3,128-128,128C57.3,256,0,198.7,0,128C0,57.3,57.3,0,128,0C198.7,0,256,57.3,256,128z};
    \fill[white] svg{M86.3,186.2H70.9V79.1h15.4v48.4V186.2z}
                 svg{M108.9,79.1h41.6c39.6,0,57,28.3,57,53.6c0,27.5-21.5,53.6-56.8,53.6h-41.8V79.1z M124.3,172.4h24.5c34.9,0,42.9-26.5,42.9-39.7c0-21.5-13.7-39.7-43.7-39.7h-23.7V172.4z}
                 svg{M88.7,56.8c0,5.5-4.5,10.1-10.1,10.1c-5.6,0-10.1-4.6-10.1-10.1c0-5.6,4.5-10.1,10.1-10.1C84.2,46.7,88.7,51.3,88.7,56.8z};
  }
}

\newcommand\orcidicon[1]{\href{https://orcid.org/#1}{\mbox{\scalerel*{
\begin{tikzpicture}[yscale=-1,transform shape]
\pic{orcidlogo};
\end{tikzpicture}
}{|}}}}

\usepackage{hyperref} 

\begin{document}
%

\title{Engineering Yeast Cells to Facilitate Information Exchange}
%
%
%

\author{Nikolaos~Ntetsikas$^*$,~\IEEEmembership{Student Member,~IEEE,} 
        Styliana~Kyriakoudi$^*$,
        Antonis~Kirmizis,
        Bige Deniz Unluturk,~\IEEEmembership{Member,~IEEE,}
        Andreas Pitsillides,~\IEEEmembership{Life Senior,~IEEE,}
        Ian F. Akyildiz,~\IEEEmembership{Life Fellow,~IEEE} and
        Marios Lestas,~\IEEEmembership{Member,~IEEE} \\$^*$\normalsize These authors contributed equally
\thanks{N. Ntetsikas and M. Lestas are with Frederick Research Center and Frederick University, Nicosia, 1036, Cyprus (email: st023422@stud.frederick.ac.cy, eng.lm@frederick.ac.cy).}}%

\maketitle

\begin{abstract}

Although continuous advances in theoretical modelling of Molecular Communications (MC) are observed, there is still an insuperable gap between theory and experimental testbeds, especially at the microscale. In this paper, the development of the first testbed incorporating engineered yeast cells is reported. Different from the existing literature, eukaryotic yeast cells are considered for both the sender and the receiver, with $\alpha$-factor molecules facilitating the information transfer. The use of such cells is motivated mainly by the well understood biological mechanism of yeast mating, together with their genetic amenability. In addition, recent advances in yeast biosensing establish yeast as a suitable detector and a neat interface to in-body sensor networks. The system under consideration is presented first, and the mathematical models of the underlying biological processes leading to an end-to-end (E2E) system are given. The experimental setup is then described and used to obtain experimental results which validate the developed mathematical models. Beyond that, the ability of the system to effectively generate output pulses in response to repeated stimuli is demonstrated, reporting one event per two hours. However, fast RNA fluctuations indicate cell responses in less than three minutes, demonstrating the potential for much higher rates in the future.

\end{abstract}


%
\IEEEpeerreviewmaketitle

\section{Introduction}
MC deals with the application of communications theory, for the characterization of information exchange among entities which employ molecules or particles \cite{Eckford}. The main focus has been on applications related to biological systems \cite{Akyildiz POV}. In that respect, MC merges the field of communications with systems biology. In particular, it aims to apply communications principles upon the mechanisms and methods which are of molecular nature, and are employed by living or artificial entities. Applications which have been considered in the literature, include the Internet of Bio-Nano-Things (IoBNT) \cite{Akyildiz POV}, disease diagnosis \cite{Akan}, disease localization \cite{Okaie}, targeted drug delivery \cite{Chabibi} and molecular sensing \cite{Lobsiger}. It must be noted that beyond biological systems, MC has been considered to serve other applications, as for example, industrial and underwater cases \cite{Haselmayr}.  
 
Experimental setups are crucial for the MC field, as they can be harnessed to validate the plethora of theoretical findings. At the same time, they constitute a solid onset for practical applications to emerge. Despite the vast literature on mathematical tools and results, practical testbeds have so far been limited, especially at the molecular level. This deficiency can be, in many aspects attributed to the limited cooperation between electrical engineers and biologists. 

The extensive review that we present in Section \ref{testbeds}, highlights the fact that most experimental testbeds are flow based and involve artificial components at the transmitter (Tx)/receiver (Rx) ends, as well as the particles used to facilitate information exchange. MC testbeds which involve living entities, have primarily focused on the use of bacteria (prokaryotic cells) while the use of eukaryotic (such as human) cells has been limited to only two testbeds. Those involve endothelial \cite{Felicceti} and mammalian HeLa cells \cite{Nakano2}. Moreover, in the pursued mathematical models and their analysis, the cell signalling pathways which pose significant dynamic effects, have been ignored and an E2E approach is adopted instead. This can be attributed to the limited information available at the time on the underlying signaling pathways, and their mathematical characterization. Eukaryotic cells have distinct advantages over prokaryotic cells with respect to their relevance to human centered applications, and their further consideration in the MC field is thus of paramount importance.
 
In this work, for the first time in the MC literature, we use \textit{Saccharomyces cerevisiae} (yeast hereafter) as our model organism. The yeast mating pathway, is employed to investigate cell-to-cell communications from a communications theory perspective. Yeast, due to its genetic amenability and the wealth of knowledge on its operation, constitutes an ideal eukaryotic model organism for the validation of the plethora of theoretical findings in the MC field. Further, recent advances in molecular biology have established yeast as a model organism for the development of biosensors. They suggest that any analytical tools and results developed from a molecular communications perspective, can be harnessed to establish the properties (e.g. event detection rates) of such biosensing devices and promote their integration to the in-body area network system. This is expected to bring MC closer to practical applications and advance the IoBNT paradigm towards its practical realization. The genetic amenability of yeast, also suggests that it may be used as a platform for engineering nano-machines exploring exotic applications not as of now considered in literature \cite{Hofmann}.

In Section \ref{testbeds}, we present a concise review on MC experimental platforms which have appeared in the literature, highlighting the originality of our work. In Section \ref{Apps} we explain how the same platform can be used as a baseline for the development of a number of yeast biosensors, whose properties in terms of event detection rates can be characterized using the tools and methods presented in this paper. This highlights the relevance of the presented work to practical applications. In Section \ref{system description}, we present a thorough overview of the inherent functionality of our considered system,  providing a foundational understanding for subsequent analyses and discussions.  In Section \ref{system model} we present the system model, describing the underlying processes and mathematical models which are employed to characterize them. This leads to the development of an E2E model, which incorporates all elements of the considered system: transmitter, receiver and diffusive channel. The diffusive channel response poses significant challenges to obtain, and this is resolved in the case of a reaction term being absent thus rendering the solution tractable. In Section \ref{Experimental setup}, we describe the experimental setup and in Section \ref{results} we present the obtained experimental results, which highlight good agreement with the theoretical predictions thus contributing towards the validation of the theoretical models, which has been the main objective of this work. Beyond that, it is shown that Rx re-stimulation with pulse-shaped inputs of synthetic pheromone can lead to pulse-shaped outputs. Those can yield event detection rates of one event per two hours, which is rather slow. Further experimentation reveals, that this slow response can be attributed to the protein translation phase. The measured RNA levels highlight the quick up-regulation of the Rx response, showing a near-maximum up-regulation in just three minutes after stimulation. This implies that, protein translation contributes to further time delay of the system's output, and that if alternative reporting mechanisms can be employed, much higher event rates may be achieved. This will be pursued in the future.





\section{Experimental Testbeds} \label{testbeds}

The testbed presented in this paper is relevant to microscale processes. Testbeds addressing the microscale features of MC deal with phenomena such as molecular-level propagation and reactions. In contrast, macroscale MC exclusively concentrates on physical quantities and communication metrics, treating the MC signal similarly to conventional communications. This approach overlooks individual molecular phenomena, emphasizing a broader perspective on the macroscopic aspects of MC. As such, we first make a short review on macroscale testbeds, and we devote most of the discussion to microscale setups.   
 
Examples of macroscale experimental testbeds include \cite{Bhattaharjee}, \cite{McGuiness}, \cite{Yetimoglu} and \cite{Pampu}. These can be classified into air based \cite{Bhattaharjee}, \cite{McGuiness}, \cite{Pampu} and water based \cite{Yetimoglu} depending on the channel medium. The work in \cite{Bhattaharjee}, deals with molecular information exchange inside a pipe environment where a water-based fluorescein solution is sprayed inside the pipe, and is detected by a high speed camera. The authors in \cite{McGuiness} report on a testbed, where odor particles are propelled from an odor generator into a void pipe, and detected by a mass spectrometer. Both aforementioned testbeds are relevant to industrial applications. An air-based MC that can emulate pipe/tunnel/mine environments was also implemented experimentally in a  Multiple Input Single Output (MISO) scenario in \cite{Pampu}. Therein, light is used as the carrier signal and two transmitters share the same channel. Quasi-free error detection is achieved using a photodetector at the receiving end. A very recent microfluidic setup for macroscale MC can be found in \cite{Deng}, where the transmitter unit is a fabricated device with a signal shaping function, the receiving unit is an ultraviolet-visible (UV-Vis) spectrometer, the molecular signal is comprised by sodium hydroxide, and finally the propagation is facilitated by flow-assisted diffusion.


Most microscale experimental setups focus on the molecule/particle transfer mechanisms in scenarios which attempt to mimic real world applications. Those typically use engineered devices (e.g. pumps), rather than living entities serving as transmitters and receivers. For example, the recent testbed in \cite{Bartunik3}, \cite{Bartunik2}, \cite{Wicke} employs superpragmatic iron oxide nanoparticles (SPIONs), to facilitate information exchange rendering it suitable for prototyping drug delivery applications. Flow driven transport is employed with a SPION injection pump serving as the transmitter and a magnetic susceptometer used for signal detection. There, transmission rates up to 6 bits/sec have been reported. Another notable recent attempt is reported in \cite{Brand}, where a closed loop system is realized aimed at mimicking the cardiovascular system. In this testbed, the authors utilize media-based modulation of the biocompatible  Green Fluorescent Protein variant "Dreiklang" (GFPD) molecules which serve as messengers. Similar to the previous testbed, background flow is maintained by a pump with a spectrometer or spectrophotometer recording the fluorescence of the GFPD molecules once they have been turned ON by an optical (LED) transmitter.

Other flow based testbeds include \cite{Farsad}-\cite{Pan}. In \cite{Farsad}, a combination of acids and bases serve as information carriers, which are transmitted by a peristaltic pump inside flowing filtered water. These chemical signals cause a pH change, which is detected by a multi-chemical platform that converts the pH reading into an electric potential. The work in \cite{Tuccito}, further developed in \cite{Fichera}-\cite{Cali3}, present an experimental setup which harnesses fluorescent carbon quantum dots (CQDs) as information carriers, which are injected into a background fluid flow using a microvalve, and are detected by optical detectors. An in-vessel MC concept can also be found in \cite{Unterweger}, where an electronic pump injects magnetic nanoparticles inside an aqueous medium. The messenger particles propagate in the background flow which is maintained by a second pump, and are received by a susceptometer receiver. In a similar context, in \cite{Wang} and \cite{Pan} two vascular flow-inspired testbeds are presented. In \cite{Wang} three peristaltic pumps are used to transmit salt (for bit 1) and water (for bit 0) inside a narrow pipe with a background flow of water. The receiver uses an electrical conductivity reader, with the observed output being linearly related to the salt concentration. In \cite{Pan}, a peristaltic pump emulating the human heart, pumps pigment water molecules, which are detected by a receiving color sensor and then demodulated. A human glucose bio-sensor application is emulated by the testbed in \cite{Koo}. Glucose molecules are injected using a syringe inside a saline solution undergoing advection-diffusion, which are then detected by a receiving nano-chip. The aforementioned methods, employ fluid flow in tubes. Moreover, microfluidics have also been employed in recent literature. The testbed in \cite{Angerbauer}, uses salinated water as the information carrier with the salinity levels in the background flow detected using a microfluidic chip. Further, a 2x2 MIMO experimental setup that uses programmable syringe pumps as transmitters, fluorescence microscopes as receivers and fluorescent beads as messenger molecules that propagate via flow-induced diffusion, is reported in \cite{Unluturk2}.

Different from our experimental setup, the aforementioned testbeds incorporate artificial senders, information carriers and receivers. However, a number of testbeds have been reported in the literature involving living entities. Bacterial quorum sensing is harnessed in \cite{Yi Liu}, where the sender is realized using either engineered protein molecules or \textit{E.coli} inside alginate-gelatine beads. This sender "device" sends AI-2 quorum sensing molecules upon receiving the ligand "SAH", which are received by reporter \textit{E.coli} bacteria that are also assembled in alginate-gelatine beads and express fluorescence upon AI-2 induction. The propagation of the messenger molecules is facilitated via diffusion. \textit{E.coli} senders are also documented in \cite{Grebenstein} and \cite{Kirchner}, which respond to a light-emitting diode and pump proton molecules in the extracellular medium. As these diffuse towards the receiver, they change the pH of their surrounding environment, and this change is detected by a pH receiving sensor. A biocomputing platform where engineered bacteria function as logic gates, is reported in \cite{Martins}. Therein, IPTG molecules are injected into the fluidic medium and propagate towards the bacteria-based logic gates via free diffusion. The bacteria-produced ammonia and hydrogen molecules, propagate towards a receiving electrochemical sensor through a narrow tube, adhering to flow assisted diffusion. The hydrogen and ammonia molecules in turn, produce pH changes which are sensed by the receiver. In \cite{Bicen2}, bacteria are used as receivers which detect externally provided AHL molecules in a microfluidic channel via a fluorescence output. 

Different from the above testbeds, in \cite{Abbasi} neuronal cells are incorporated. An input probe generates a suitable waveform \textit{in vivo}. After the signal is received by the first neuron, the neuron  propagates its response through the axon to the rest of the neurons, and once it reaches the last one, the response is transferred through a probe to an external computer. Despite the introduction of living entities in the aforementioned testbeds, this is done partially, in the sense that artificial components are still necessary, thus compromising the bio-compatibility and the relevance to practical applications. 

Very limited works have so far considered biological entities serving as both transmitters and receivers. Bacteria-based communications that employ bacteria at both the sending side and the receiver, can be found in \cite{Terrell} and \cite{Nakano}. The authors in \cite{Terrell} present a testbed where electronic-encoded input is transduced to hydrogen peroxide via electrochemically controlled oxygen reduction. This biomolecular input, triggers bacteria cells which produce acyl homoserine lactone (AHL) and these molecules are diffused towards two other bacteria populations. One population produces $\beta$-gal molecules, which are converted to electronic signals through electrochemical oxidation and detected by an external device. The other expresses a red fluorescent protein which acts as a successful transmission confirmation. Potential applications include biological connectivity and protein recognition. Externally controlled bacteria-to-bacteria communication was also implemented experimentally in \cite{Nakano}. There, an external processing unit can stimulate a bacteria population, which can convert the electromagnetic input signal to chemical stimuli through an interface made from artificially synthesized materials (ARTs). In response to the chemical input, this bacteria population acts as a transmitter, which sends secreted fluorescent molecules to a receiving bacteria population also equipped with ARTs as a proper interface with the same external processing unit. The propagation of the fluorescent molecules is facilitated by diffusion through a gap junction channel between the two bacteria populations. Gap junctions are intercellular channels formed by specialized proteins, that create a small path between cytosolic environments of adjacent cells. Potential applications of \cite{Nakano} include pattern formation which can be useful for tissue regeneration.

The aforementioned testbeds involve mostly bacteria whose prokaryotic nature can be limiting in terms of their applicability in human-centered applications. Eukaryotic cells offer distinct advantages with respect to their biocompatibility,  however their consideration is limited to the works of  \cite{Felicceti} and much earlier in \cite{Nakano2}. In \cite{Felicceti}, blood platelets were stimulated using Thrombin to secrete CD40L messenger molecules, which employ free diffusion in a trans well setup to reach human umbilical vein endothelial cells (HUVECs). Upon reception by the eukaryotic HUVECs, the output gene VCAM-1 was measured by flow cytometry. The setup aims to mimic the frequent communication of platelets and endothelial cells inside human blood vessels. A micro-platform for the characterization of inter-cellular communication was reported much earlier in \cite{Nakano2} employing mammalian HeLa cells. Communication between these mammalian cells is triggered by external stimulation, and is facilitated by gap junction channels through which calcium molecules diffuse from one cell to the adjacent one. Signal detection is facilitated by fluorescence microscopy. The testbed relies on exogenous intervention as the gap junction channels are formed on predefined paths established using lithography.

\section{Yeast Biosensing}\label{Apps}
The yeast based testbed presented in this paper can be used to support basic research activities in MC, validating existing theoretical findings or using it as a baseline for developing new modeling approaches and methods to be exploited by the MC research community in the future. However, one important aspect that needs to be addressed is the type of biomedical applications that could be supported from this technology.
One example is the use of yeast to develop logic gates for biological circuits \cite{Hofmann}. However, an application which is highly relevant to the developed testbed is yeast biosensing, much due to the flexibility that yeast cells present, when it comes to cell and genome engineering. As indicated in the recent review of \cite{Wahid} most yeast-based biosensors are still in the early developmental stage, with only a few prototypes tested for real applications. As such, the testbed described herein can contribute towards this direction. MC theory can supplement the ongoing efforts by characterizing and optimizing the event detection rates and the interfacing/communication with the in-body area network that will facilitate the transfer of the sensed readings to external monitoring systems. Such interfacing can be offered by the fluorescent output of the considered system which can be integrated with recently developed smart pills/capsules \cite{Ding} to support digital twin applications, or the realization of the IoBNT.                  

There already have been some efforts in modifying yeast cells for environment sensing; scenarios include both human-centered cases as well as industrial settings \cite{Wahid}. Some examples of substances that can be sensed by yeast include acetic acid, copper, genotoxins, methanol and androgens among others. It is estimated, that around one hundred yeast biosensors have been developed at a laboratory stage, which showcases the potential that yeast has as a sensing nanomachine, and implies the major role that it will have in both industry and healthcare in the future. 

Apart from the still-developing yeast biosensors, there are indeed a few prototypes that have been reported. In \cite{Ostrov}, on-site pathogen surveillance was achieved by introducing a lycopene production pathway to \textit{S. cerevisiae} cells, and embedding those to a dipstick that can contact the liquid under test. Another yeast sensing prototype that also resulted in a commercial product, describes Yestrosens \cite{Lobsiger}, a \textit{S. cerevisiae} biosensor which is field-portable and storable, and can detect endocrine-disrupting chemicals. The prototype is unique, in that it is capable of sending the read-out signals to a smartphone, something that is crucial for IoBNT applications. 

\begin{figure}[!hbt]
\centering
\includegraphics[width=\columnwidth]{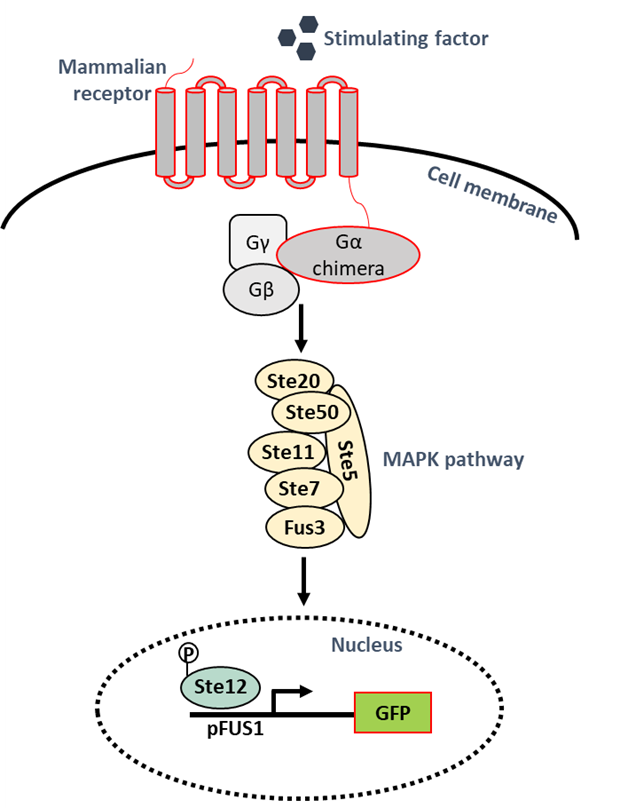}
\caption{Schematic diagram of the yeast mating signaling pathway coupled to the heterologous expression of a mammalian GPCR.\\The GPCR occupancy by the stimulating ligand on the cell surface triggers the splitting of the receptor-bound heterotrimeric G-protein into the G$\alpha$ subunit, and the G$\beta$$\gamma$ dimer. Replacement of the endogenous G$\alpha$ subunit by a chimeric yeast/mammalian G$\alpha$ protein couples the heterologous receptor to the endogenous MAPK signaling pathway and the downstream expression of the mating-responsive GFP reporter gene. The engineered elements of the pathway are outlined in red color.}
\label{app}
\end{figure}

As part of our work in progress, we briefly outline how the presented testbed can be used as the basis for the development of a number of yeast biosensors. The detailed study and implementation of the envisioned system will be part of a future work. We propose a yeast biosensor made from \textit{S. cerevisiae}  cells for human health applications. The model of the yeast mating system, can be further harnessed to offer a practical experimental platform towards the investigation of molecular signaling, in an extended range of mammalian receptors. To this end, G protein-coupled receptors (GPCRs) of mammalian origin that share common features with the classical GPCR signal transduction pathway, could be heterologously expressed in yeast to couple the native pheromone communication system. The employment of engineered yeast strains expressing mammalian GPCRs, constitutes a robust strategy for the detection of biomarkers linked to human diseases, drug discovery and pharmaceutical screening. In \textit{S. cerevisiae}, the GPCRs act through heterotrimeric G proteins composed of G$\alpha$, G$\beta$ and G$\gamma$ subunits, which are held together under non-stimulated conditions. Upon activation of the GPCRs by their cognate pheromones, the receptors induce the exchange of GDP with GTP on the G$\alpha$ subunit, which is eventually freed from the G$\beta \gamma$ dimer. The dissociated subunits, regulate the activity of intracellular effector proteins of the Mitogen Activated Kinase (MAPK) pathway that propagates the signal for cell conjugation. The efficient interaction of the heterologous receptors with the effectors of the yeast pheromone mating pathway, can be achieved by replacing the endogenous G$\alpha$ subunit of the heterotrimeric complex with a yeast/mammalian chimeric G$\alpha$, that retains efficient binding with the yeast G$\beta \gamma$ dimer. In particular, the extreme C-terminal region of the yeast G$\alpha$ can be replaced with residues of mammalian G$\alpha$ subunits, enabling the G$\alpha$-chimeras to functionally couple a wide range of mammalian receptors to the yeast signalling machinery, as illustrated in Fig.\ref{app}. Accordingly, the expression of the endogenous receptor that is encoded by the \textit{STE2} gene can be eliminated through targeted \textit{STE2} gene disruption to enable correct trafficking, of the heterologous receptor to the cell surface.

\section{System Description} \label{system description}
The system under consideration is based on the yeast mating pheromone response pathway \cite{Lestas}. The budding yeast \textit{S. cerevisiae} exists in one of the two mating types: MAT$\alpha$ and MATa. Both secrete distinct peptide pheromones, that they use to detect mating partners of the opposite sex type nearby. Specifically, MATa cells secrete a-factor pheromone, a 12 amino-acid long peptide, while MAT$\alpha$ cells secrete $\alpha$-factor pheromone, a 13 residue peptide. The secreted pheromones are perceived by surface receptors on yeast cells of the opposite mating type.
 
\begin{figure}[!hbt]
\centering
\includegraphics[width=\columnwidth]{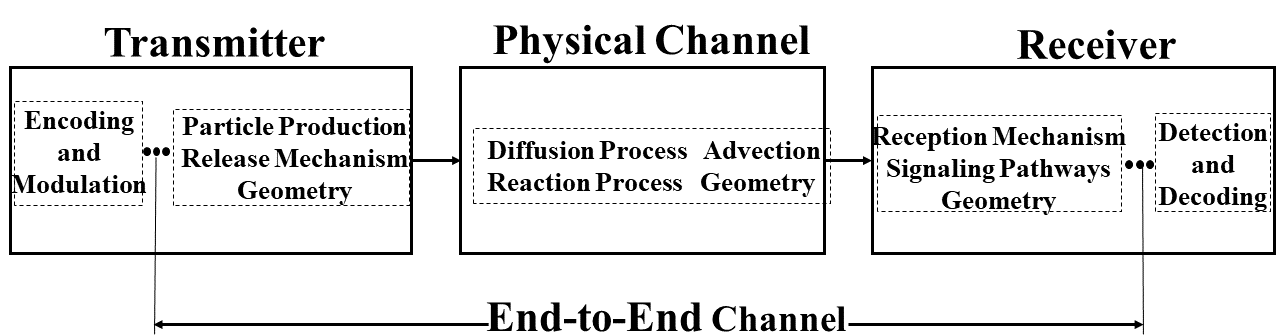}
\caption{End-to-End system.}
\label{E2E}
\end{figure}

\begin{figure}[!hbt]
\centering
\includegraphics[width=\columnwidth]{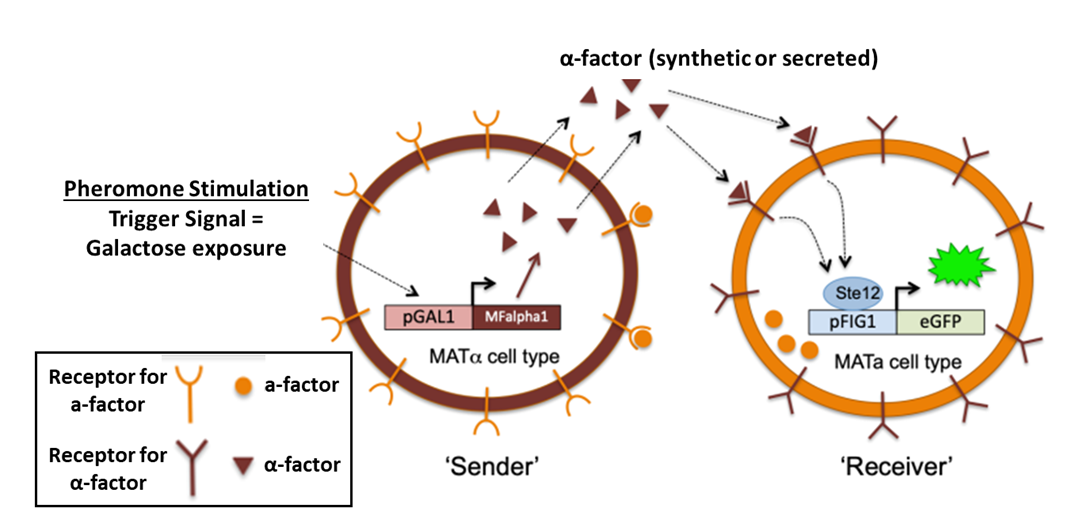}
\caption{Fundamental point-to-point yeast communication system.}
\label{System Description}
\end{figure}

Leveraging the response of MATa cells when stimulated by the $\alpha$-factor pheromone molecules produced by the MAT$\alpha$ senders, we proceed on designing our theoretical system model together with our testbed. From a modelling point of view, we construct an overall E2E model, which we divide in three main parts as in Fig.\ref{E2E}: the Tx (Transmitter) subsystem, the channel subsystem, and the Rx (Receiver) subsystem. We consider a fundamental point-to-point scenario, where our system essentially comprises of one transmitting unit, one receiving unit and a fluid channel in between. The transmitter and the receiver are represented by a MAT$\alpha$ and a MATa cell respectively (haploid cells), as shown in Fig.\ref{System Description}. The MAT$\alpha$ cell, can be engineered so as to respond to galactose triggers, and secrete $\alpha$-factor molecules through appropriately coupling the internal galactose transduction pathway with the existing mechanism that participates in the production of the $\alpha$-factor peptide molecules. Those $\alpha$-factor particles, constitute the pheromone which is ready to be exported in the fluidic medium, and is suitable for yeast culture growth.

\begin{figure}[!hbt]
\centering
\includegraphics[width=\columnwidth]{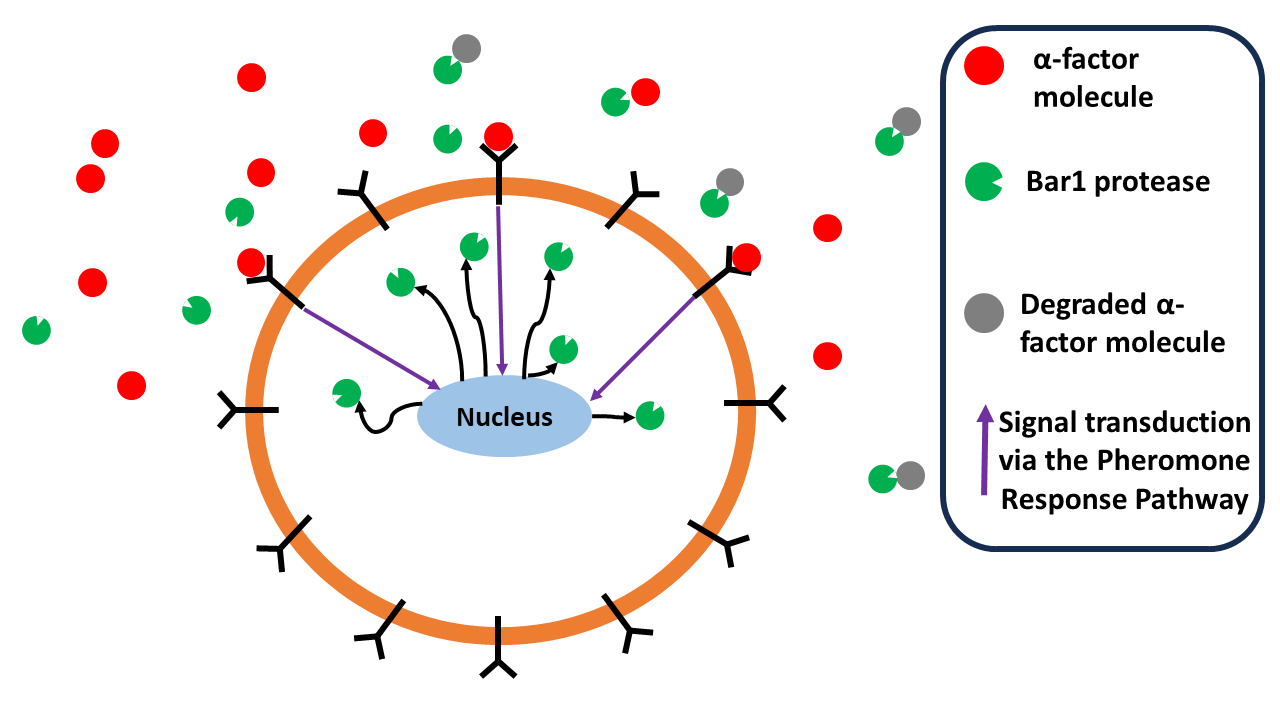}
\caption{Bar1 protease degrading pheromone particles.}
\label{Bar1}
\end{figure}

The pheromone propagates towards the receiver via diffusion, where the individual particles follow Brownian motion in three dimensional space. When the MATa cell is stimulated by pheromone secreted by a nearby cell of the opposite mating type (MAT$\alpha$), it undergoes a number of changes in order to prepare for mating. These changes include altered expression of a couple hundred genes, cell-cycle arrest, morphological alterations like polarised growth towards the mating partner, and finally fusion of the two partner cells to form a MATa/MAT$\alpha$ diploid cell \cite{Bardwell}. One of the transcribed genes, referred to in the literature as \textit{BAR1} \cite{Barkai}, encodes for a protease that diffuses outside of the MATa cell and induces degradation to the incoming pheromone, as shown in Fig.\ref{Bar1}.
In our approach, we assume that the fluidic medium is static, hence the propagation mechanisms involved are reaction and diffusion, with no advection or other flow phenomena involved. The galactose and pheromone pathways, are the two signal transduction mechanisms that will be considered here. The former characterizes the $\alpha$-factor production, while the latter quantifies the mating gene output. 

\section{System Model}\label{system model}
The experimental setup comprises of three main components: the transmitter cells (wild type MAT$\alpha$ cells which secrete $\alpha$-factor particles), the diffusion channel (“type of liquid” in which the secreted $\alpha$-factor particles diffuse to the receiver cells) and the receiver cells (MATa cells which express green fluorescence upon the binding of $\alpha$-factor to their receptors). We refer to our previous work in \cite{Ntetsikas} for more details. Each part of the model is associated with different dynamics, as a result of the signaling pathways involved as well as the diffusion process. In this section, we provide mathematical models which can be used to characterize the three components of the considered system. The objective has been in delivering a computational representation of the E2E model, towards the development of a simulation platform which can be used by the research community as a cost-effective tool, for pursuing further investigations on this model organism without the need to resort to practical experimentation. The experimental setup considered in this study incorporates those three sub-blocks, and the aim has been to validate the theoretical findings with the experimental results. This will play a major role in building confidence on the suitability of the theoretical model, to accurately represent the actual system. The developed simulation platform thus, holds the capability to emulate practical experiments serving as a pivotal asset in our study. For the most part, our models include differential equations that describe the temporal dynamics of the various genes and proteins involved, whose concentrations are depicted in the corresponding equations in Section \ref{Transmitter} and Section \ref{Receiver} within square brackets for ease of readability.


\subsubsection{Transmitter}\label{Transmitter}
In \cite{Ntetsikas}, exogenously provided pheromone inflow into the fluidic medium was achieved by direct injection as a first step, to report on a receiver model suitable for our system. In this work, this approach is extended with a wild type MAT$\alpha$ cell (the opposite sex type and mating partner of the MATa cell) serving as the sending unit. MAT$\alpha$ cells are well known to respond to galactose stimulation, when available in the medium by inducing the expression of galactose metabolic (GAL) genes \cite{Mitre}, \cite{Bhat}.

In the mathematical analysis presented in this section, the underlying biological processes that dictate the production of GAL genes after galactose stimulation, are presented. Although various modeling approaches exist in the literature, in this work we adopt coupled differential equations to characterize the temporal dynamics of the most relevant protein and RNA molecules involved in the GAL pathway. The engineered generation of $\alpha$-factor molecules under the GAL pathway is viewed as a controlled process, where galactose acts as the triggering signal. At the theoretical level, we consider that the promoter of the Galactokinase 1 gene, otherwise known as \textit{GAL1}, can be fused with the \textit{MFalpha1} coding region which encodes the $\alpha$-factor peptide. Although the experimental development of controlled $\alpha$-factor production upon galactose stimulation remains a future task, the mathematical equivalent of this process is provided here as an extension of the GAL pathway model. The generation of the $\alpha$-factor peptide molecules that are ready to be exported, can be described by simple ODE models that include first order generation and degradation terms \cite{Klipp2}. 

Following the analysis in \cite{Mitre}, the galactose pathway is divided in two main parts: the metabolic pathway, and the genetic pathway. The metabolic branch consists mainly of two main functions. The first one is the transportation of the extracellular galactose to the intracellular space by the transporter protein Gal2 (denoted as $G_2$), via  carrier-facilitated diffusion. This acts as a positive feedback loop of the intracellular galactose concentration. The transport function is dependent mainly on the balance between intracellular ($G_i$) and extracellular ($G_e$) galactose concentration, and is given by the following equation \cite{Mitre}:
 \begin{IEEEeqnarray}{rCl}
T(G_2,G_i) = r_{max} [G_2]\Bigl(\frac{[G_e]}{K+[G_e]}-\frac{[G_i]}{K+[G_i]}\Bigr)
\label{Transport}
\end{IEEEeqnarray}
with $r_{max}$ representing the maximum possible transport rate that can be achieved, and K is the concentration needed so that the reaction rate between $G_2$ and $G_e$ as well as $G_2$ and $G_i$, reaches its half-maximum value. Thus, K can also be defined as the half-maximum rate for \eqref{Transport}. Essentially, K denotes the "affinity" that $G_e$ and $G_i$ have towards the transporter protein $G_2$. After imported in the cell, galactose is converted into a phosphorylated form represented by $G_p$, by the protein kinase $Gal1p$, however tagging the intracellular galactose for further degradation due to the phosphorylation process. Therefore, this process eventually acts as a negative feedback loop term. The authors in \cite{Mitre} represent this phosphorylation process using a Michael-Menten kinetics model, with $\sigma$ being the maximum phosphorylation rate which can be achieved by the protein $Gal1p$. The half-maximum activation $\kappa_p$ denotes the $Gal1p$ concentration needed to achieve the half-maximum phosphorylation rate. The Michaelis-Menten kinetics parameters are given by:

\begin{flalign}
&\sigma(G_i) = \frac{\kappa_{GK}K_{IU}K_{IC}}{K_{IU}K_m+K_{IC}[G_i]} &\\
&\kappa_p(G_i) = \frac{(K_m+[G_i])K_{IU}K_{IC}}{K_{IU}K_m+K_{IC}[G_i]} &
\end{flalign}

The experimentally measured phosphorylation rate of $G_i$ is expressed by $\kappa_{GK}$, with the constants $K_{IC}$ and $K_{IU}$ representing the "competitive inhibition constant" and "non competitive inhibition constant" correspondingly, and $K_m$ being another phosphorylation constant obtained from the literature. The resulting phosphorylated form of galactose, is further metabolized inside the MAT$\alpha$ cell, having a metabolizing rate $\delta$. Apart from being metabolized, $G_p$ is also diluted within the MAT$\alpha$ cell having a rate $\mu_\alpha$. This dilution term corresponds to the case where there is no glucose inside the medium. As such, the time dynamics of the $G_p$ concentration can be described by the following differential equation:


 \begin{IEEEeqnarray}{rCl}
\frac{d[G_p]}{dt} = \frac{\sigma(G_i)}{\kappa_p([G_i])+[G_p]}[G_1][G_i]-(\delta+\mu_\alpha)[G_p]
\label{Gp}
\end{IEEEeqnarray}

The metabolic reactions are characterized by much faster dynamics compared to the dynamics of the proteins involved in the genetic branch of the galactose pathway (in the order of hundreds). This implies that Quasi-Steady State (QSS) analysis can be applied in \eqref{Gp}, considering only the steady state of $G_p$. This simplification enables the derivation of a system of ODEs, describing the genetic branch of the galactose pathway. As we consider an environment where galactose and glucose co-exist and given the fact that glucose is a repressor of galactose, we proceed with the analysis in the fashion of \cite{Mitre}. The probability of galactose transportation across the transmembrane protein in the presence of glucose in the medium, is described by the scaling factor $y(R)$:
  \begin{IEEEeqnarray}{rCl}
y(R) = (1-y_b) + \frac{y_b}{y_c+R}
\label{yR}
\end{IEEEeqnarray}
where $1-y_b$ is the basal probability of galactose transport when glucose is absent, $y_c$ is the half-maximum transport repression by glucose and R represents the glucose concentration inside the the medium. The rate of galactose repression when glucose is present, is dictated by the Hill function:
\begin{IEEEeqnarray}{rCl}
x(R) = \frac{1}{(\frac{R}{x_c})^{n_x}+1}
\label{xR}
\end{IEEEeqnarray}
where $n_x\geq 1$, and $x_c$ is the half-maximum of the repressive process.

The genetic branch, deals with the transcription and translation of the GAL genes and the corresponding proteins that result from the process. The mRNAs that are transcripted upon galactose induction include the \textit{GAL3}, \textit{GAL80}, \textit{GAL2} and \textit{GAL1}, herein also denoted as $M_3$, $M_{80}$, $M_2$ and $M_1$ correspondingly. The transcription factors which act at the promoter level, are the corresponding proteins. The proteins Gal1p and Gal3p activate the galactose network by enabling the transcription of the GAL genes when galactose is present in the medium. In contrast, Gal80p acts as a repressor and inhibits the transcription when galactose is absent, by binding on the promoters of the GAL genes. As soon as galactose enters the medium, part of the transcribed and then translated Gal3p is phosphorylated by the intracellular galactose, thus reaching an activated state. The activated form of Gal3p binds to the Gal80p dimers and cancels their inhibitory function, thus enabling the further transcription of the GAL genes. This in turn results in increased levels of Gal3 and Gal1 protein concentration. Part of the Gal1p is also phosphorylated by the intracellular galactose, which then replaces the activated Gal3p dimers in the GAL promoters. This leads to an even more efficient continuous transcription of GAL genes inside the cell. For the representation of the aforementioned mechanisms, models for the dynamics of \textit{GAL80}, \textit{GAL3}, \textit{GAL1} and $G_i$ genes and their corresponding proteins are sought. We adopt the modeling approach of \cite{Mitre} to consider a system of ODEs that describe the time evolution of these compounds. The reader is referred to \cite{Mitre}, for a detailed description of the GAL pathway dynamics, as well as the derivation of the ODE system describing them. Below, we show the set of ODEs that we consider in our analysis:

\begin{flalign}
&\frac{d[M_3]}{dt} = \kappa_{tr,3}x(R)R_1([G_{80}],[G_3],[G_1],[G_i])-\nonumber\\&(\gamma_{M,3}+\mu(R))[M_3] \label{M3}&\\
&\frac{d[M_{80}]}{dt} = \kappa_{tr,80}x(R)R_1([G_{80}],[G_3],[G_1],[G_i])-\nonumber\\&(\gamma_{M,80}+\mu(R))[M_{80}] &\\ 
&\frac{d[M_2]}{dt} = \kappa_{tr,2}x(R)R_2([G_{80}],[G_3],[G_1],[G_i])-\nonumber\\&(\gamma_{M,2}+\mu(R))[M_2] &\\
&\frac{d[M_1]}{dt} = \kappa_{tr,1}x(R)R_4([G_{80}],[G_3],[G_1],[G_i])-\nonumber\\&(\gamma_{M,1}+\mu(R))[M_1] \label{GAL1}&\\ 
&\frac{d[G_3]}{dt} = \kappa_{tl,3}[G_3]-\Bigl(\gamma_{G,3}+\mu(R)+\frac{\kappa_{C,3}[G_i]}{K_S+[G_i]}\Bigr)[G_3] \label{G3}&\\
&\frac{d[G_{80}]}{dt} = \kappa_{tl,80}[G_{80}]-(\gamma_{G,80}+\mu(R))[G_{80}] &\\
&\frac{d[G_2]}{dt} = \kappa_{tl,2}[G_2]-(\gamma_{G,2}+\mu(R))[G_2] &\\
&\frac{d[G_1]}{dt} = \kappa_{tl,1}[G_1]-\Bigl(\gamma_{G,1}+\mu(R)+\frac{\kappa_{C,1}[G_i]}{K_S+[G_i]}\Bigr)[G_1] \label{G1}&\\
&\frac{d[G_i]}{dt} = r_{max} y(R)[G_2]\Bigl(\frac{[G_e]}{K+[G_e]}-\frac{[G_i]}{K+[G_i]}\Bigr)-\nonumber\\ &\frac{2\sigma (G_i)[G_1]}{k_p(G_i)+\sqrt{k_p[G_i]^2+\frac{4\sigma (G_i)[G_i][G_1]}{\delta}}}-\nonumber\\ &[G_i]\Bigl(\frac{\kappa_{C,3}[G_i]}{K_S+[G_i]}+\frac{\kappa_{C,1}[G_i]}{K_S+[G_i]}\Bigr)-\mu(R)[G_i] \label{Gi}&
\end{flalign}


In the above system of ODEs, the transcription rates of $M_3$, $M_{80}$, $M_2$ and $M_1$ are mentioned as $\kappa_{tr,3}$, $\kappa_{tr,80}$, $\kappa_{tr,2}$ and $\kappa_{tr,1}$, while their protein translation rates are denoted by $\kappa_{tl,3}$, $\kappa_{tl,80}$, $\kappa_{tl,2}$ and $\kappa_{tl,1}$. The negative feedback loop term in every ODE from \eqref{M3} to \eqref{Gi}, includes one degradation and one dilution term. The RNA and protein degradation terms are denoted as $\gamma_{M,3}$, $\gamma_{M,80}$, $\gamma_{M,2}$, $\gamma_{M,1}$, $\gamma_{G,3}$, $\gamma_{G,80}$, $\gamma_{G,2}$ and $\gamma_{G,1}$. Due to the presence of glucose, the total dilution rate $\mu(R)$ is increased compared to the case where glucose is absent, and it is dependent on the glucose concentration R. The terms $\kappa_{C,3}$ and $\kappa_{C,1}$ in \eqref{G3} and \eqref{G1}, represent the rate of $G_3$ and $G_1$ activation by the intracellular galactose $G_i$. $K_S$ is the intracellular galactose concentration that results in half-maximum activation rate for both $G_1$ and $G_3$. The probability that the promoter of one of the GAL genes is occupied by a transcription factor at a given time so that the transcription occurs, is represented by the fractional transcription $R_n$. 

\begin{flalign}
&\Omega = 1+\sum_{k=1}^{n} \Bigl(\frac{K_{80}}{[G_{80}]}\Bigr)^{2k}+\sum_{k=1}^{n} \Bigl(\frac{[G_3][G_i]}{K_3(K_S+[G_i])}\Bigr)^{2k}+\nonumber\\&\sum_{k=1}^{n} \Bigl(\frac{[G_1][G_i]}{K_1(K_S+[G_i])}\Bigr)^{2k}\\
&R_n([G_{80}],[G_3],[G_1],[G_i]) =1-\frac{1}{\Omega} \label{Rn}
\end{flalign}
The subscript n, denotes the number of transcription factor binding sites (Upstream Activating Sequences, $[UAS]_g$) with the constants $K_1$, $K_3$ and $K_{80}$ given by:

\begin{flalign}
&K_1 = \frac{\sqrt{K_{D,1}K_{B,3}K_{B,1}}(\gamma_{G,1}+\mu_\alpha)}{\kappa_{c,1}}& \\
&K_3 = \frac{\sqrt{K_{D,3}K_{B,3}K_{B,3}}(\gamma_{G,3}+\mu_\alpha)}{\kappa_{c,3}}& \\
&K_{80} = \sqrt{K_{D,80}K_{B,80}}&
\end{flalign}

The dissociation constant for the dimerization of the GAL genes before binding to a $[UAS]_g$, is represented by $K_{D,i}$ while $K_{B,i}$ denotes the dissociation constant of the binding reaction of those dimers to a $[UAS]_g$. Both constants can be obtained from the existing literature. We complement the above system of ODEs, using appropriate models for pheromone generation within the MAT$\alpha$ cell. Specifically, the transcription of the $\alpha$-factor gene from the GAL1 promoter, is dictated by the presence of the transcription factor Gal4p which induces the GAL1 promoter inside the MAT$\alpha$ cell. Due to the very close interconnection between the \textit{GAL1} gene and the $\alpha$-factor, the transcription process of the latter can be described by the following ODE which resembles \eqref{GAL1}, as we consider Gal4p to be the transcription factor of the $\alpha$-factor gene within the cell's nucleus:

\begin{flalign}
&\frac{d[MFalpha1]}{dt} =\kappa_{tr,1}x(R)R_3([G_{80}],[G_3],[G_i])-\nonumber \\&k_{deg}[MFalpha1]
\label{MFalpha}
\end{flalign}
with $k_{deg}$ denoting the degradation rate of the $\alpha$-factor mRNA concentration. Finally, the dynamics of the $\alpha$-factor protein synthesis and peptide processing (which is to be exported in the extracellular medium), can be described by the following ODEs:
\begin{flalign}
\frac{d[\alpha_p]}{dt}= k_{tr,\alpha}[MFalpha1]- k_{degP}[\alpha_p]\label{alphaPr}& \\ 
\frac{d[\alpha_{pep}]}{dt}= k_{pep,\alpha}[\alpha_p]-k_{degPep}[\alpha_{pep}] \label{alphaPep}& 
\end{flalign}
where $k_{tr,\alpha}$ represents the translation rate of the $\alpha$-factor pheromone mRNA molecules, $k_{degP}$ the degradation rate of the $\alpha$-factor protein form, $k_{pep,\alpha}$ the peptide processing rate within the MAT$\alpha$ cell and $k_{degPep}$ the $\alpha$-factor peptide degradation rate. The reader can refer to \cite{Klipp2} for more information regarding equations \eqref{MFalpha}-\eqref{alphaPep}.

\subsubsection{Diffusion}\label{Diffusion}

After being secreted to the extracellular medium in its peptide form, the $\alpha$-factor molecules undergo both reaction and diffusion processes at the microscale. Those dictate the pheromone concentration across the propagation medium over time. These processes associated with particle Brownian motion, cause the molecules to propagate towards the Rx through the fluidic channel. The relevant Reaction-Diffusion (RD) equation is documented in \cite{Giese}:

 \begin{IEEEeqnarray}{rCl}
\frac{\partial \alpha}{\partial t} = D_\alpha\laplacian{\alpha}-k_{re}B\alpha-k_\alpha\alpha + S_\alpha
\label{pheromone RD}
\end{IEEEeqnarray}

where, $\alpha=\alpha(x,y,z,t)$ denotes the $\alpha$-factor concentration, measured in $moles/m^3$, or $M$, while $k_\alpha$ characterizes a small autodegradation phenomenon resulting from the biochemical properties of the $\alpha$-factor molecules within the fluidic medium, and is measured in $1/sec$. Here it is treated as a constant. The diffusion coefficient is denoted by $D_\alpha$ and it is measured in $m^2/sec$. The $S_\alpha=S_\alpha(x,y,z,t)$ term represents the pheromone concentration flow into the channel (pheromone generation term), either by the transmitter (MAT$\alpha$ cell), or by direct pheromone injection \cite{Ntetsikas}, and is measured in M/sec. Moreover, the concentration term $B=B(x,y,z,t)$ denotes the Bar1 \cite{Giese} concentration. The reaction rate $k_{re}$ is measured in $1/(M\cdot sec)$ and represents the interaction rate between the $\alpha$-factor and Bar1 molecules. To render the analysis simpler, it is also treated as a constant.  Bar1 diffuses in the medium, but in the opposite direction of the pheromone molecules \cite{Giese}, undergoing degradation processes similar to the $\alpha$-factor. The associated RD equation for Bar1 is given by:
\begin{equation}
\frac{\partial B}{\partial t} = D_B\nabla^2{B}-k_BB+S_B
\label{Bar1 RD}
\end{equation}
Where $k_B$ represents an autodegradation rate similar to $k_\alpha$, and $S_B$ corresponds to the Bar1 generation term, similar to $S_\alpha$. The term $D_B$ corresponds to the diffusion coefficient of the Bar1 molecules. The solution of the system of coupled Partial Differential Equations (PDEs) \eqref{pheromone RD} and \eqref{Bar1 RD}, is an open problem which we aim to address in the near future. The complexity associated with the solution implies that one may resort to numerical approximations for its characterization. In this work, we ignore the Bar1 effects from \eqref{pheromone RD}, and by setting B=0 in \eqref{pheromone RD}, we consider the following RD equation:

\begin{IEEEeqnarray}{rCl}
\frac{\partial \alpha}{\partial t} = D_\alpha\laplacian{\alpha}-k_\alpha\alpha + S_\alpha
\label{pheromone RD2}
\end{IEEEeqnarray}

We can obtain the impulse response of the propagation channel by utilizing the Green's Function Theory, to find an analytical expression for $\alpha$. Our initial step is to transpose our problem to solve for a function $\hat{L}G\left(\mathcal{S}\middle| \mathcal{S}^\prime\right)=\delta^3\left(\mathcal{S}-\mathcal{S}^\prime\right)=\delta\left(x-x^\prime\right)\delta\left(y-y^\prime\right)\delta\left(z-z^\prime\right)$, where $\mathcal{S}=\left(x,y,z\right)$ represents an observation point in space, and $\mathcal{S'}=\left(x',y',z'\right)$ represents the impulse source point. By applying the three-dimensional (for x,y,z) spatial Fourier transform to \eqref{pheromone RD2} we obtain the following PDE:

\begin{equation}
\frac{\partial \Tilde{\alpha}(k,l,v,t)}{\partial t} =( -D_{\alpha}g^2-k_{\alpha})\tilde{\alpha}(k,l,v,t),
\label{Fourier RD}
\end{equation}
where, $g^2=k^2+l^2+v^2$, with k, l and v being the counterparts for the spatial variables x, y and z in the Fourier domain. Solving the above PDE for $\alpha$ yields:

\begin{equation}
\Tilde{\alpha}(k,l,v,t) = \tilde{\alpha}(k,l,v,0)e^{(-D_{\alpha}g^2-k_{\alpha})t}
\label{Fourier solution}
\end{equation}

The initial condition for the $\alpha$-factor corresponds to an impulse of amplitude  $a_0$ at  t=0, namely, $\alpha\left(x,y,z,0\right)=a_0\delta^3\left(\mathcal{S}-\mathcal{S^\prime}\right)$. Also adopting the notation $X = kx+ly+vz$ and $X^\prime=kx^\prime+ly^\prime+vz^\prime$ for convenience, we obtain the Inverse Fourier Transform of \eqref{Fourier solution} as:

\begin{equation}
\begin{aligned}
 &\alpha = \alpha_0(\frac{1}{2\pi})^3\int^\infty_{-\infty} \int^\infty_{-\infty}\int^\infty_{-\infty}e^{iX}e^{-iX'}e^{-D_{\alpha}g^2t}e^{-k_\alpha t} \,dk\,dl\,dv 
   \\ &=\frac{\alpha_0}{\sqrt{4 \pi D_{\alpha}t}^3}e^{-(\frac{(x-x')^2+(y-y')^2+(z-z')^2}{4 \pi D_{\alpha}t}+k_{\alpha}t)} = \\ &\frac{\alpha_0}{(4 \pi D_{\alpha}t)^{3/2}}e^{-(\frac{r_{Rx}^2}{4 \pi D_{\alpha}t}+k_{\alpha}t)} &
\end{aligned}
\label{CIR}
\end{equation}
where $r_{Rx}$ represents the Rx distance from the pheromone transmission point. For the general case where there is no impulse input for t=0, $F^{-1}\left(\widetilde{a}\left(k,l,v,0\right)\right)=\phi\left(x, y, z\right)$ corresponds to the initial  $\alpha$-factor particle distribution inside the medium. In that case, \eqref{CIR} needs to be solved again by substituting the term $e^{-iX'}$ with the three-dimensional spatial Fourier transform of $\phi\left(x, y, z\right)$. The result of \eqref{CIR}, serves as the input to our receiver system whose mathematical representation is detailed below based on our previous work in \cite{Ntetsikas}.

\subsubsection{Receiver}\label{Receiver}
In this subsection we describe our receiver model. Based on the existing literature for the initiation of the pheromone response, here we outline the process during which the MATa cell translates the input pheromone to the output gene expression. The \textit{FUS1} response is of interest, and comprises the considered Rx output within the developed model. The $\alpha$-factor particles that bind to the receiver's receptors, activate the corresponding receptor protein named Ste2, which subsequently elicits the initiation the MAPK signaling cascade. The MAPK pathway is responsible to orchestrate essential processes for sexual reproduction and begins with the dissociation of the heterotrimeric G protein resulting in the $G_{\alpha,GTP}$ compound and the $G\beta \gamma$ dimer. The latter, binds to the scaffold complex, which is formed by the scaffold protein Ste5, and the bound kinases Ste11, Ste7 and Fus3. Ste5 plays a major role in signaling pathway specificity, as it accumulates the bound kinases in a specific area of the cell, and it protects those from auto/hetero-inhibition and degradation. Ste5 is essential for the phosphorylation of Ste11 upon mating pheromone stimulation.

The latter part of this cycle which is of great concern (also known as the G-cycle), is the pheromone-dependent binding of the protein Ste20 to the already formed complex. Such binding of Ste20 with Ste5 and $G\alpha \beta \gamma$, triggers its phosphorylation, leading to the phosphorylation of Ste11 through direct interaction. Upon phosphorylation of the Ste11, Ste7 gets activated through binding with Ste11. Ultimately, the signaling cascade activates the Fus3 protein which then dissociates from the scaffold and stimulates downstream mating responses. It is important to note, that the aforementioned individual chemical compounds that take part in the G-cycle, already exist in the cell prior to stimulation. Furthermore, during the mating response of \textit{S. cerevisiae} a large number of genes are upregulated, followed by cell cycle arrest and morphological changes. Within that context, the transcription of the \textit{FUS1} mating gene is induced by the transcription factor known as Ste12 \cite{Klipp},\cite{Houser},\cite{Bardwell}.

Ste12 resides within the nucleus of the MATa cell, where two different protein complexes are located. The first complex, consists of two proteins known as Dig1 and Dig2. These act as Ste12 repressors, inhibiting the activation of Ste12, but also protecting it from being degraded over the course of time. Once free from Dig1 and Dig2 repressors, Ste12 forms Ste12 homodimers, which then bind to specific sites on DNA promoters, known as the Pheromone Response Elements (PREs) \cite{Houser}, thus initiating the transcription of the \textit{FUS1} gene. The protein Fus1 belongs to the Pheromone Response Pathway, and contributes to the actual mating process of every pair of haploid yeast cells. The second complex, contributes to another pathway known as the Filamentous Growth Pathway. Filamentous growth, is a pheromone-regulated growth response in which the Tec1
transcription factor binds to filamentous growth promoters. By doing so, it stimulates the expression of genes that allow the bringing of the spatial gap between MATa and MAT$\alpha$ cells. In steady state conditions and prior to pheromone reception, two tripartites consisting of Ste12, Dig1 and Dig2, as well as Ste12, Dig1 and Tec1 are formed inside the MATa cell.

After pheromone stimulation, Fus3 phosphorylates the Dig1-Ste12-Dig2 and Dig1-Ste12-Tec1 complexes, facilitating the loosening of the bond between those three in each complex. After dissociating from Dig1 and Dig2, Ste12 and Tec1 form two complexes, namely one homodimer (Ste12-Ste12) and one heterodimer (Ste12-Tec1). Each of these complexes which bind to the corresponding DNA promoters, lead to the trancsription of either the mating genes (such as the \textit{FUS1}), or the filamentous growth genes. Fundamentally, Tec1 is an antagonist of the mating process, as it has been shown to bind to Ste12, undermining the overall transcription of the \textit{FUS1} mating gene and promoting filamentous growth. This antagonistic behavior of the filamentous pathway, creates a "crosstalk" when pheromone binds to the receptors of the MATa cell \cite{Houser}. Both individual pathways co-exist inside MATa cells. Although we do account for the filamentous growth antagonist in the model, the filamentous genes are omitted, with the exception of the filamentous growth transcription factor Tec1.

In \cite{Klipp}, the Pheromone Response Pathway is developed from a set of complex chemical reactions between compounds which become relevant after pheromone sensing. These reactions are represented by a set of ODEs, some of which are non-linear. However, the focus in \cite{Klipp} has been on modeling the response of specific protein complexes that contribute to the MATa cell shmoo tip formation, and orientation towards the mating partner, which is the MAT$\alpha$ cell. Actual mating genes as for example \textit{FUS1}, which, as indicated above, are of interest to our system, were neglected. To counter for this deficiency, we resort to the work of \cite{Houser} which models the \textit{FUS1} output gene (in this work depicted in \eqref{FUS1 gene}, in Appendix \ref{Appendix Rx}) related processes using an ODE approach, as in \cite{Klipp}. Therein, the authors provide a system description that includes both pathways, namely for mating and filamentous growth. Despite the significance of the models, the authors only include the processes that take place at the Rx nucleus with the MAPK activity (the phosphorylation of Fus3 which is critical for the activation of Ste12) not  taken into account, but rather modelled as a simple time-varying function. To provide adequate Fus1 modelling, we couple the approaches in \cite{Klipp} and \cite{Houser}, complementing the missing parts of the one with the details of the other and incorporating modifications where necessary. This allows the development of a comprehensive mathematical tool that can utilize the pheromone concentration at the receptor level to provide a prediction for the time evolution of Fus1. This enables interfacing our receiver model with the channel model described in Section \ref{Diffusion}. The full set of ODEs together with the model parameters are presented in Appendix \ref{Appendix Rx}.

\section{Experimental Setup} \label{Experimental setup}
To render the mathematical models of Section \ref{system model} reliable for the computational representation of the system, those need to be validated using experimental data. The inherent yeast mating mechanism and the flexibility when it comes to cell engineering, creates the opportunity for a wide range of experimental studies when using yeast as the model organism. Different mutant yeast cells can be created from modifying the yeast’s functionalities. The multifunctional capabilities of this eukaryote, can be used to study and optimize MC through altering and testing yeast cells under certain, yet modifiable conditions, with the aid of an experimental testbed. We thus develop and utilize a practical setup, which can fulfil this objective and at the same time help us draw further useful conclusions for the system under consideration. In this section, we describe the developed experimental testbed.

\subsection{System Description} \label{Experimental system description}
As previously stated, the testbed design was based on the extensively studied biological process of \textit{S. cerevisiae} mating response, that employs MAT$\alpha$ cells as biological pheromone transmitters and MATa cells for biological receivers. The latter, indicate the pheromone detection by emitting green fluorescence which is detected by our testbed equipment (Fig. \ref{System Description}).


In preparation for conjugation, pheromonal signals received by the yeast, activate the G-protein-coupled receptors on MATa and MAT$\alpha$ cells. The activation of those receptors, induce the transduction of the MAPK cascade signaling leading to activation of the mating gene-specific transcription factor Ste12. Ste12 has been shown to act as a positive regulator of approximately 3\% of the yeast genome that is involved in the mating process \cite{Dolan}, \cite{Song}. Among the upregulated genes, \textit{FUS1} is considered an early target of the pheromone response pathway as it displays at least 10-fold elevated levels of expression after exposure of MATa cells to $\alpha$-factor \cite{McCaffrey}, \cite{Trueheart}.

As a proof of principle, in this study we have engineered MATa cells that express a reporter protein called superfolder Green Fluorescence Protein (sfGFP) under the intrinsic \textit{FUS1} promoter by simultaneous deletion of the endogenous \textit{FUS1} gene, to eliminate potential fusion events with neighboring partners. Furthermore, to extend our in depth analysis of molecular communications, we have also engineered a \textit{bar1$\Delta$} deletion mutant strain to exploit its pheromone-sensitive phenotype conferred by the lack of the Bar1 protease, that cleaves and inactivates $\alpha$-factor from the surroundings of the cell \cite{Barkai}, \cite{Manney}, \cite{MacKay}.

\subsection{Materials and Methods}
\subsubsection{Yeast strains and culture conditions}
All yeast strains that were genetically engineered in the framework of this study were originated from \textit{S. cerevisiae} BY4741 (wild type, genotype MATa, ura3$\Delta$0, leu2$\Delta$0, his3$\Delta$1, met15$\Delta$0) and transformed with PCR products by the lithium acetate method \cite{Gietz}. The fus1$\Delta$::sfGFP (Bar1p+) reporter strain was constructed by genetic replacement of the \textit{FUS1} gene via homologous recombination of a PCR product containing a superfolder GFP (sfGFP) fused to a CLN2 PEST sequence and the natMX cassette for selection. The PCR product was amplified from the Gal1pORF-sfGFPdeg-natMX strain \cite{Bheda}. For the fus1$\Delta$::sfGFP \textit{bar1$\Delta$}::kanMX (\textit{bar1$\Delta$}) strain, genetic deletion of the \textit{BAR1} gene was achieved by PCR-mediated homologous recombination of the \textit{G418} drug resistance gene obtained from plasmid pE2‐Crimson-degron-kanMX (gifted from Prof. Michael Knop). All oligonucleotides used for PCR-amplifications are listed in Table \ref{sequences}.

For the co-cultivation experiments, the \textit{S. cerevisiae} BY4742 strain (wild type, genotype MAT$\alpha$, ura3$\Delta$0, leu2$\Delta$0, his3$\Delta$1, met15$\Delta$0) was employed serving as a constitutive sender of $\alpha$-factor pheromone. Yeast cells were cultured to saturation overnight at 30 °C in selective YPAD medium (0.12 g/L adenine hemisulfate salt, 10 g/L yeast extract, 20 g/L peptone and 20 g/L glucose) supplemented with appropriate antibiotics. The next day, cells were diluted to OD600 = 0.2 and grown for a few hours until OD600 reached approximately 0.6, prior to experimental induction.

\subsubsection{$\alpha$-factor induction and cell harvesting}
Yeast cells at the logarithmically growing phase (OD600 = 0.6) were stimulated with 10 $\mu$M of synthetic $\alpha$-factor (Zymo Research Corp) and samples of cell suspension were harvested at specific timepoints following initial stimulation.
For the RNA assay, aliquots of $\approx$ 3 x $10^6$ cells were harvested per each timepoint, washed with sterile water and stored at -80 °C before proceeding with RNA isolation. For fluorometric measurements, stimulated cells were adjusted to the exponential phase of growth, washed once with sterile water and aliquots of $\approx$ 3 x $10^6$ cells were transferred to a 96-well microplate (Thermo Scientific) to monitor fluorescence yield. Green emission was determined in arbitrary units (AU) using the Infinite F200 fluorescence microplate reader (Tecan Trading Ltd, Switzerland) with excitation, 485 ⁄ 20 nm; emission, 510 ⁄ 20 nm; manual gain, 100. The presented data from the fluorescence assays were obtained from at least three independent experiments.

\subsubsection{RNA isolation, reverse transcription and quantitative PCR}
Total RNA was isolated from frozen samples using the hot phenol method \cite{Schmitt} and cDNA synthesis was performed with the PrimeScript RT Reagent Kit (Perfect Real Time, TaKaRa Bio). Quantitative real-time PCR was performed on a CFX96 thermal cycler (BioRad) using the KAPA SYBR FAST qPCR Master Mix (KAPA Biosystems). ACTb ($\beta$-actin) was used as housekeeping gene and all oligonucleotide primers used in the assay are listed in Table \ref{sequences}. Relative changes in gene expression were analyzed using the 2(-Delta Delta C[T]) algorithm. All samples and standards were run in triplicates and the presented results were obtained from two independent experiments.

\begin{table}[!t]
\caption{Oligonucleotide primers used in the study}
\label{sequences}
\centering
\setlength\tabcolsep{1.5pt}
\begin{tabular}{|c||c|}
\hline
A. For PCR amplification &\\
\hline
Yeast Strain & Primer pair\\
\hline
\textit{fus1$\Delta$}::sfGFP & 
\begin{tabular}{c} F: 5’- CCTTTAAGAGCAGGATAT\\AAGCCATCAAGTTTCTGAAAAT\\CAAAatgtccaagggtgaagagct -3’\\ \hline R: 5’- CAGAATTATAGGTATA\\GATTAAATGCGAACGTCAATA\\TTATTTTCAcagtatagcgaccagcattc-3’
\end{tabular}\\
\hline
\textit{fus1$\Delta$}::sfGFP \textit{bar1$\Delta$}::kanMX &
\begin{tabular}{c} F: 5’- CGCCTAAAATCATACCAAA\\ATAAAAAGAGTGTCTAGAAGGGT\\CATATAgacatggaggcccagaatac -3’\\ \hline R: 5’- CTATATATTTGATATTT\\ATATGCTATAAAGAAATTGTACT\\CCA GATTTCcagtatagcgaccagcattc -3’
\end{tabular}\\
\hline
B. For quantitative PCR &\\
\hline
Gene & Primer pair\\
\hline
sfGFP &
\begin{tabular}{c} F: 5’- CCATTTTGGTAGAACTGGAC \\-3’\\ \hline R: 5’- CATATGGTCTGGGTATCTTG \\-3’
\end{tabular}\\
\hline
$\beta$-actin &
\begin{tabular}{c} F: 5’- CATCTTCCATGAAGGTCAAG \\-3’\\ \hline R: 5’- CTTGTGGTGAACGATAGATG \\-3’
\end{tabular}\\
\hline
\end{tabular}
\end{table}

\section{Results}\label{results}
In this section, we showcase results that stem from both experimental and computational work aiming to validate the theoretical models and extract relevant useful interpretations. Theoretical analysis includes single pheromone stimulation of the Rx cell, whilst the conducted experiments incorporate both, single and multiple stimulations. The theoretical findings are then compared against experimental results, which are obtained using engineered yeast cells stimulated by $\alpha$-factor molecules. The $\alpha$-factor molecules are either secreted by wild type MAT$\alpha$ cells which act as senders, or synthetic $\alpha$-factor is directly injected into the medium, in predetermined quantities.

For the receiver side, two types were considered, namely wild type, otherwise called as "Bar1p+" and "Bar1 knockout", otherwise abbreviated as "\textit{bar1$\Delta$}". The former yeast strain incorporates the protease "Bar1" (thus being in its natural wild type form), while the latter lacks Bar1 due to deletion of its gene. The \textit{bar1$\Delta$} receiver better matches the setting considered in the mathematical model, which renders the solution of the diffusion equation tractable. The wild type receiver, due to the secretion of the Bar1 repressor of $\alpha$-factor pheromone, is known to exhibit more responsive and robust behavior. This is reflected in its ability to generate well regulated pulses of the system output, with the gene expression being measured using the available experimental setup.

In this context, we sought to unravel whether our receiver yeast strains retain any capacity of transcriptional re-induction, a biological phenomenon of major significance for the sustained expression of certain genes, upon repeated exposures of the cell to specific stimuli \cite{Bheda}. We established an experimental design to determine the transcriptional re-induction capacity of the \textit{GFP} gene in Bar1p+ and \textit{bar1$\Delta$} strains by quantifying GFP fluorescence intensities over time following repeated exposures to stimulating pheromones. In our approach, we performed a three-pulse stimulation (i1, i2 and i3) of bulk cells with 10 $\mu$M of $\alpha$-factor that lasted for 1 minute on intermittent basis, followed by a 2 hour-removal of the stimulating factor amid the inductions to promote signal repression (r1, r2 and r3). The MATa cell cultures that were used, were physically grown in the available wet lab.

In the experiments where single pheromone stimuli was realized, both the expression of
the RNA transcripts of the \textit{GFP} gene, as well as the protein that results from RNA translation were investigated. By characterizing both stages of the receiving procedure, we were able to determine the dynamics of the reception process in search of methodologies with which to achieve faster responses.




\subsection{RNA levels of GFP}\label{RNA}
The first set of experiments, involved MATa cells excited with 10$\mu$M of synthetic $\alpha$-factor. We consider both types of MATa cells, namely the Bar1p+ and \textit{bar1$\Delta$} strains.

\begin{figure}[htpb]
  \centering
  \subfloat[Elevated RNA levels of GFP following induction of Bar1p+ cells with synthetic $\alpha$-factor.]{\includegraphics[width=3in]{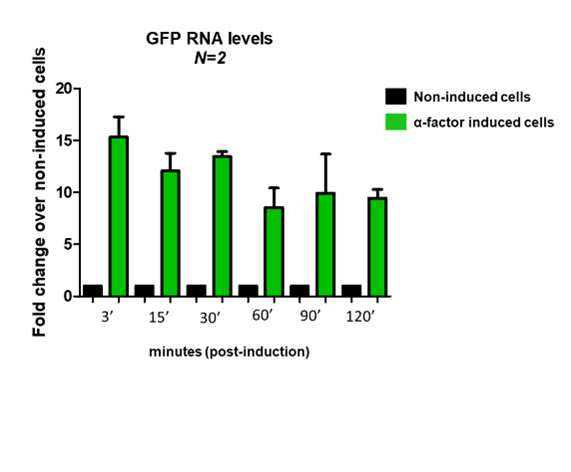}%
 \label{single Bar1 RNA}}
  \hfill
  \subfloat[Elevated RNA levels of GFP following induction of \textit{bar1$\Delta$} cells with synthetic $\alpha$-factor.]{\includegraphics[width=3in]{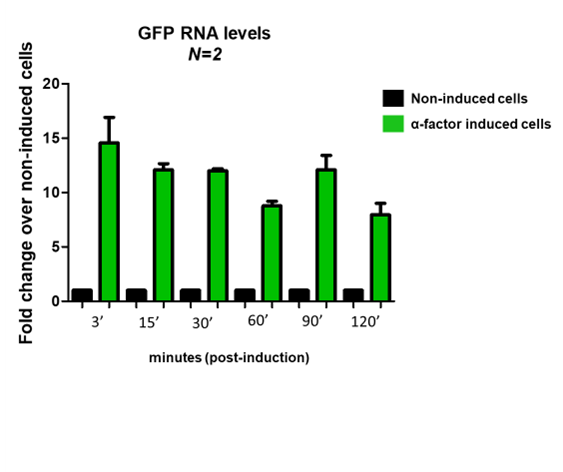}%
 \label{single no Bar1 RNA}}
  \caption{Stimulation of MATa cells with 10 $\mu$M synthetic $\alpha$-factor and assessment of GFP RNA levels.\\GFP transcript levels were determined by quantitative PCR following the induction of Bar1p+ (Bar1 present) and \textit{bar1$\Delta$} (Bar1 Knockout) cells with 10 $\mu$M of exogenously provided $\alpha$-factor. RNA levels were monitored for 120’ following initial induction, at the indicated timepoints. The values were normalized to the expression of $\beta$-actin and presented as a fold change over non-induced cells. Values from two independent experiments in both Fig.\ref{single Bar1 RNA} and Fig.\ref{single no Bar1 RNA}, are presented as mean $\pm$ standard error of the mean (SEM).}
  \label{Single Stimulation RNA}
\end{figure}

The first thing to note in Fig.\ref{single Bar1 RNA} and Fig.\ref{single no Bar1 RNA}, is that in both yeast strains the RNA up-regulation reaches its peak within 3 minutes after pheromone reception, which indicates that cell response is relatively fast. This quick response of both yeast strains implies that, fast detection rates are feasible upon the availability of real-time detection tools for the RNA response. 

Real-time RNA sequence detection has been reported in the literature \cite{Guet}-\cite{Kukhtevich}. These works indicate that genetically encoded RNA aptamers that specifically bind to an exogenously supplied fluorophore, have emerged as a promising tool for the precise tracking and quantitation of transcripts in living cells. One such aptamer is the so called ‘Spinach’, that has been shown to successfully label RNA molecules in living \textit{S. cerevisiae} cells. In particular, the ‘Spinach’ sequence can be cloned between the coding region and the 3’-UTR of any target gene and elicit green fluorescence, following the binding of the GFP chromophore analogue, named DFHBI, to the aptamer. In parallel, innovative imaging technologies have been developed to provide spatiotemporal and quantitative characterization of both high and low abundant transcripts in real time. This holds immense significance, as it signifies the potential to harness the rapid RNA response of yeast cells to attain elevated event rates. 

By further analyzing the graphical summaries, we observe similar levels of GFP transcription in both Bar1p+ and \textit{bar1$\Delta$} cells, which likely suggests that hypersensitive mutants depleted of the Bar1 protease show decreased accuracy at high concentrations of $\alpha$-factor. This observation is presumably attributed to the inability of these cells to detect pheromone gradients, especially when reaching a saturated state, due to excessive concentrations of synthetic $\alpha$-factor that were used for induction. Our observation comes in line with the observation in \cite{Klipp}, where it is mentioned that stimulating a MATa cell with an $\alpha$-factor concentration of 10$\mu$M, can lead to cell saturation. This phenomenon is caused by the fact that yeast cells can only increase the up regulation of the genes involved in the Pheromone Response Pathway up to a degree. Namely, excessive increase of the stimulation ”strength” will not result in a likewise amplification of the output response, but will put the receiver cell in a saturated state.

\subsection{Protein abundance of the GFP response}\label{Protein}
\subsubsection{Single $\alpha$-factor stimulation}\label{Single protein}
Beyond measuring the RNA levels of the GFP transcript as an output, here we also present its protein form, which is the result of the corresponding RNA translation process that takes place after transcription. This serves as the main indicator of the communication, mainly due to the already existing detection equipment that can record protein fluorescence in real time. Following a similar approach as in Section \ref{RNA}, we conduct experiments using both, namely the Bar1p+ and the \textit{bar1$\Delta$} strains following a single-minute induction with 10$\mu$M of $\alpha$-factor. 
 
\begin{figure}[htpb]
  \centering
  \subfloat[Enhanced green fluorescent protein expression of Bar1p+ cells following induction with exogenously provided $\alpha$-factor.]{\includegraphics[width=2.5in]{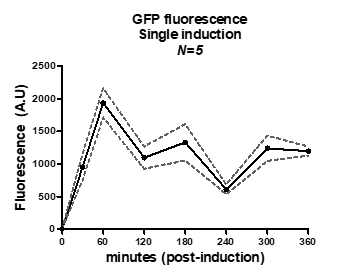}%
 \label{single Bar1 protein}}
  \hfill
  \subfloat[Enhanced green fluorescent protein expression of \textit{bar1$\Delta$} cells following induction with exogenously provided $\alpha$-factor.]{\includegraphics[width=2.5in]{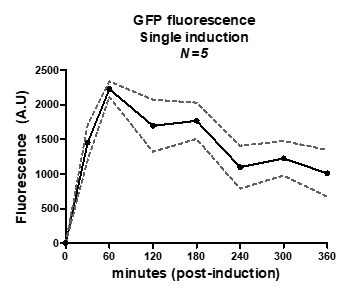}%
 \label{single no Bar1 protein}}
  \caption{Stimulation of MATa cells with 10 $\mu$M synthetic $\alpha$-factor, and quantification of GFP fluorescence.\\GFP fluorescence was monitored in both receiver strains at regular intervals (as indicated on the graphs) for a total timespan of 6h after initial stimulation with 10 $\mu$M of synthetic $\alpha$-factor whose exposure to the cells was for 1 minute. Fluorescence units from each point were normalized against the fluorescence emitted by cells not treated with $\alpha$-factor. Values from five independent experiments in both Fig.\ref{single Bar1 protein} and Fig.\ref{single no Bar1 protein}, are presented as mean $\pm$ SEM. The solid line denotes the GFP fluorescence expressed in arbitrary units (A.U) and the dotted line represents the error bars calculated at each data point.}
  \label{Single Stimulation Protein}
\end{figure}

The GFP response presented in Fig.\ref{Single Stimulation Protein}, reveals that for both strains a significant increase of green fluorescent signal, was detectable on a fluorescence reader from the early time point of 30 minutes. Fluorescence peak was observed at 60 minutes after induction for both yeast strains. This indicates that the \textit{bar1$\Delta$} strain, even though lacking the Bar1 protease, does not express hypersensitivity for the case where the stimulation lasts for as little as 1 minute similarly to what has been observed with the RNA as output. Comparing the protein response to that of the RNA response, one can observe that the overall protein response needs about 30 minutes to reach at a substantial level, indicating that the protein form of the GFP response is considerably slower. This is attributed to the nature of the RNA translation process, which introduces further time delay in the Rx response. The already existing-although sophisticated techniques for RNA detection, stand in stark contrast to conventional GFP tracking techniques, which are more readily accessible to typical wet labs. The prospect of leveraging the swift RNA response, through RNA fluorophores, opens up new ways for accelerating the detection and monitoring of events, presenting a valuable alternative to the more commonplace methodologies relying on GFP tracking.

\subsubsection{Multiple $\alpha$-factor stimulations} \label{Pulses}
In the following set of experiments, we consider multiple consecutive pheromone stimulation inputs, aiming to investigate the ability of the receiver to generate corresponding distinct output pulses which can be used for effective detection of the input generative events. Towards this goal, we consider the case of three consecutive pulses of synthetic $\alpha$-factor excitation having duration of 1 minute each. Against this background, each pulse was followed by the complete removal of the stimulating factor for 120 minutes to promote signal repression. This resembles the on-off keying (OOK) modulation scheme in traditional communication systems. Both yeast strains were utilized with each pulse induced by an injection of 10 $\mu$M of $\alpha$-factor concentration.

\begin{figure}[htpb]
  \centering
  \subfloat[Maintained re-induction capacity of the GFP gene in Bar1p+ cells following repeated stimulations with synthetic $\alpha$-factor.]{\includegraphics[width=3in]{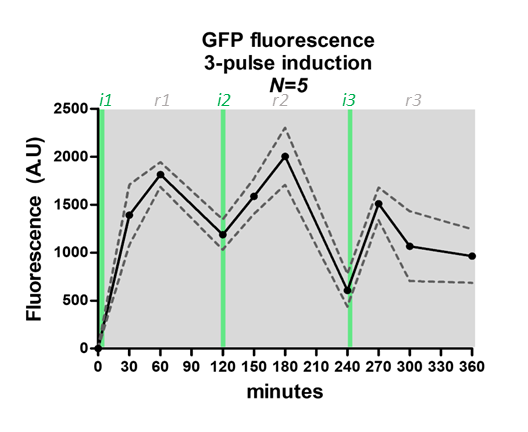}%
 \label{restim Bar1 protein}}
  \hfill
  \subfloat[Decreased re-induction capacity of the GFP gene in \textit{bar1$\Delta$} cells following repeated stimulations with synthetic $\alpha$-factor.]{\includegraphics[width=3in]{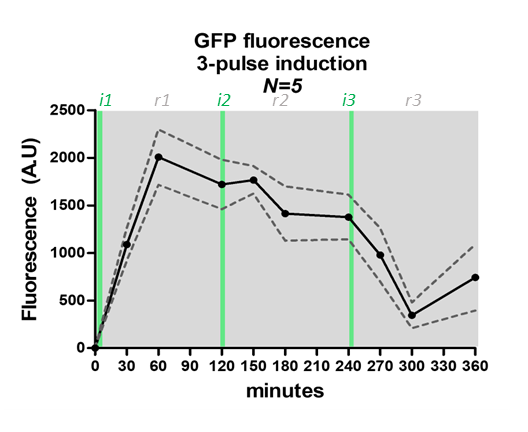}%
 \label{restim no Bar1 protein}}
  \caption{Three consecutive stimulations with 10 $\mu$M synthetic $\alpha$-factor and assessment of GFP re-induction capability in MATa cells.\\GFP fluorescence was monitored in Bar1p+ and \textit{bar1$\Delta$} cells subjected in repeated $\alpha$-factor inductions (green shaded areas - i1, i2 and i3) and respective signal repressions (grey shaded areas - r1, r2 and r3). Pheromonal inductions were performed at the indicated timepoints and maintained for 1 minute while signal repressions endured for 120 minutes. Fluorescence units from each point were normalized against the fluorescence emitted by cells not treated with $\alpha$-factor. Values from five independent experiments in both Fig.\ref{restim Bar1 protein} and Fig.\ref{restim no Bar1 protein} are presented as mean ± SEM. The solid line denotes the GFP fluorescence expressed in arbitrary units (A.U) and the dotted line represents the error bars calculated at each data point.}
  \label{Restimulation Protein}
\end{figure}

The recorded GFP responses of the two MATa cell strains are shown in Fig.\ref{Restimulation Protein}. We observe that the Bar1p+ yeast type is able to generate distinct output pulses whereas the \textit{bar1$\Delta$} strain fails to achieve that, implying restimulation deficiency. The capacity of the Bar1p+ strain to generate distinct pulse-shaped outputs, may correlate to the integral capability of the \textit{FUS1} promoter to respond to successive stimulations, likely being more resistant to pheromonal saturation effects.

Plotting the GFP intensities for both strains, reveals that Bar1p+ cells demonstrate somehow more steady high levels of GFP yield throughout the induction rounds (i1, i2 and i3), while \textit{bar1$\Delta$} cells present a dramatic decrease of GFP signal in i3 and slightly smaller in i2, as compared to i1. Collectively, these data demonstrate that the transcriptional capacity of the \textit{FUS1} gene is maintained in the Bar1p+ strain whilst it is dramatically impaired in \textit{bar1$\Delta$} cells, suggesting that lack of the Bar1 protease and impaired gradient-sensing of $\alpha$-factor signaling, exert a detrimental effect on the re-induction capability of a mating-responsive gene.

\subsection{System Validation}
In this subsection, we showcase the potential of a simulating platform that similarly with the experimental testbed, can characterize yeast response to pheromone stimuli, thus creating an opportunity for virtual experiments that complement the physical ones. As a fundamental first step towards this end, the mathematical model described in Section \ref{system model}, is validated with the experimental results presented in Section \ref{Protein} and \ref{RNA}. The model combines the RD equation \eqref{pheromone RD2} which characterizes the pheromone propagation, and equations \eqref{alpha factor ODE}-\eqref{FUS1 ODE} which quantifies the Fus1 protein abundance given an input pheromone concentration, akin to the way that our experiments quantify the GFP response of the yeast strains that we used. 
The reception model is presented in detail, in Appendix \ref{Appendix Rx}, where the ODEs that describe the Pheromone Response Pathway are provided, together with some useful definitions. More information about the mathematical model can be found in the PheroMolCom project website https://pheromolcom.frederick.ac.cy/.

\begin{figure}[htpb]
  \centering
  \subfloat[Comparison of the experimental and theoretical results for the single stimulation case. RNA levels are reported.]{\includegraphics[width=2.7in]{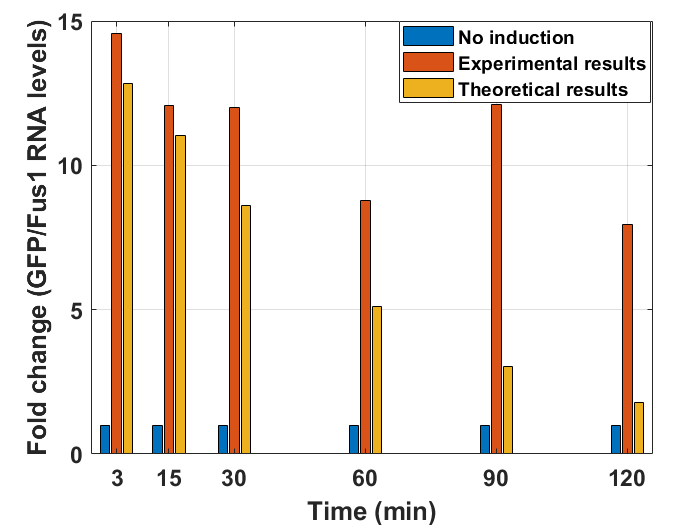}%
  \label{validation1}}
  \hfill
  \subfloat[Comparison of the experimental and theoretical results for the single stimulation case. Protein levels are reported.]{\includegraphics[width=2.5in]{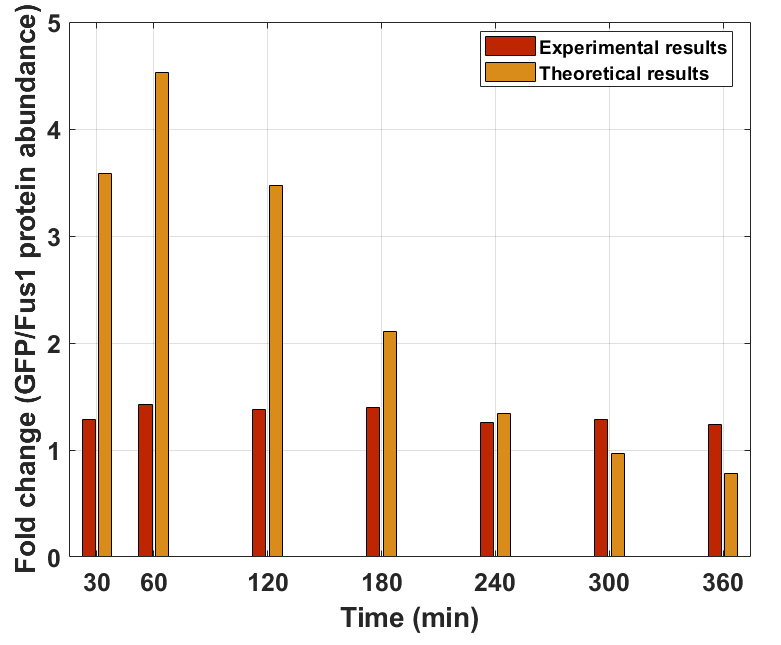}%
 \label{validation2}}
  \caption{Validation graphs for experimental/theoretical result cross matching in the \textit{bar1$\Delta$} mutant cell.}
  \label{validation}
\end{figure}

We consider the scenario which includes a single pheromone stimulation, where \textit{bar1$\Delta$} cells are excited by exogenously provided synthetic $\alpha$-factor. Both the RNA and protein form of the \textit{FUS1} and \textit{GFP} genes are used as indicators of the communication output, of the computational model and experiments correspondingly. In the computational setting, 10 $\mu$M of $\alpha$-factor are applied and the impulse response of the channel (derived in Section \ref{Diffusion}) is considered as the input to the computational model. We juxtapose experimental outcomes with computational predictions across various time points, following the application of the $\alpha$-factor input. Such comparative analysis is presented in Fig. \ref{validation}. The analytical model demonstrates a satisfying concordance, successfully forecasting the time instance at which the peak of the Rx output manifests. Beyond merely capturing the apex response of the Rx cell, the model is able to track the temporal evolution pattern of the GFP output in both RNA and protein forms.

However, discernible disparities surface at the amplitude level, attributable to stochastic influences such as noise phenomena, imparting a degree of unpredictability to the experiments. These unforeseen perturbations may emanate from molecular propagation intricacies or within the pheromone response pathway of the Rx cells \cite{Kaern}. Notably, the proliferation of yeast cells during the growth phase introduces an element of asynchrony in cell cycles within the yeast cell culture. This asynchrony implies divergent responses among cells at the time of stimulation, yielding variations in the GFP output among individual cells. Finally, background fluorescence of yeast cells during measurements must also be taken into account, to provide more accurate predictions for the fold change of the amplitude levels. It is imperative to note that our model remains deterministic in nature, as the incorporation of noise processes into our modeling methodology is reserved for future exploration.

The outcomes depicted in Fig. \ref{validation} are indicative of promising strides toward establishing a computational platform. This platform holds the potential to inform the design of future experiments with a level of fidelity, conducive to advancing our understanding of the underlying dynamics in yeast communication.

\subsection{Yeast Cells as Senders}

\begin{figure}[htpb]
  \centering
  \subfloat[Transcriptional induction of GFP in co-cultures of Bar1p+ cells with $\alpha$-factor-secreting wild type MAT$\alpha$ cells.]{\includegraphics[width=3in]{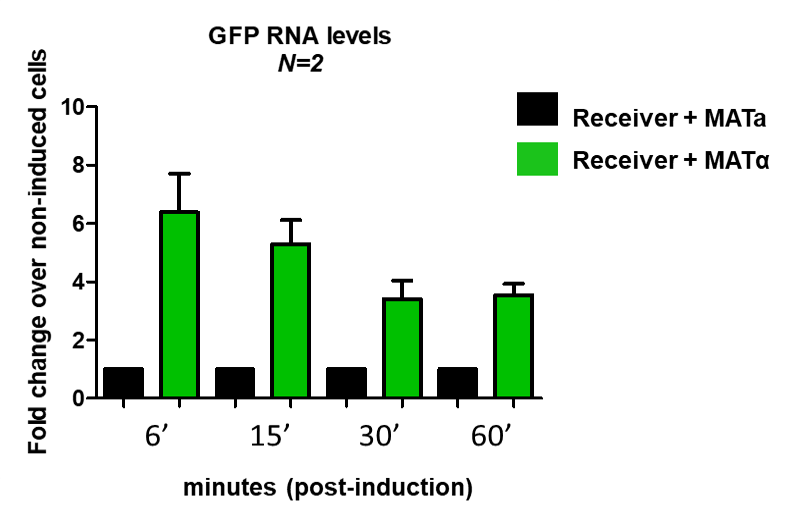}%
 \label{noBar1 co-cultures}}
  \hfill
  \subfloat[Transcriptional induction of GFP in co-cultures of \textit{$bar1\Delta$} cells with $\alpha$-factor-secreting wild type MAT$\alpha$ cells.]{\includegraphics[width=3in]{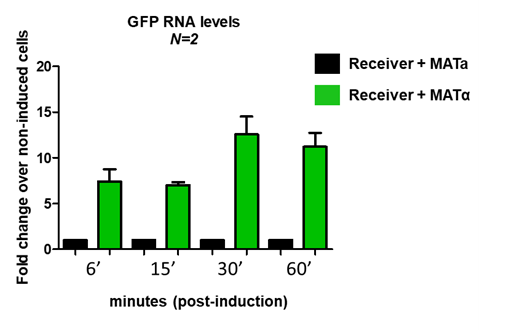}%
 \label{Bar1 co-cultures}}
  \caption{Transcriptional induction of GFP in co-cultures of Bar1p+ or \textit{bar1$\Delta$} cells with $\alpha$-factor-secreting MAT$\alpha$ cells (senders).\\ GFP RNA levels were determined by quantitative PCR following the co-cultivation of Bar1p+ (Fig.\ref{noBar1 co-cultures}) or \textit{bar1$\Delta$} cells (Fig.\ref{Bar1 co-cultures}) with $\alpha$-factor-secreting wild type MAT$\alpha$ cells (at 1:1 ratio). RNA levels were quantified at 6’, 15’, 30’ and 60’ following the mixing of sender and receiver cells in the culture vessel. The values were normalized to the expression of $\beta$-actin and presented as a fold change over the receiver cells that were co-cultured with wild type MATa cells (no induction). Values from two independent experiments in both Fig.\ref{noBar1 co-cultures} and Fig.\ref{Bar1 co-cultures}, are presented as mean ± SEM.}
  \label{yeast co_cultures}
\end{figure}

Conducting experiments with yeast cells serving as pheromone senders, involved interactions between wild type MAT$\alpha$ cells with constitutive $\alpha$-factor expression, and MATa receivers. We aim to scrutinize the physiological response of the MATa cells within the natural yeast cell-to-cell communication framework (see Fig. \ref{System Description}). At the receiver side, we consider two cases; the first one corresponds to the use of Bar1p+ Rx cells; The second case corresponds to the use of \textit{$bar1\Delta$} Rx cells. For both cases, we maintained the same MAT$\alpha$-MATa cell number ratio, namely 1:1. The RNA transcripts of the GFP response at different time points in both MATa strains, is plotted in Fig.\ref{yeast co_cultures}. Both responses depict fast RNA dynamics, which is in line with some of the observations in Section \ref{RNA}. 

Nevertheless, the overall response of both receivers in this natural setting of induction present altered dynamics compared to the transcriptional response reported in Section \ref{RNA}. This observation is ascribed to the fact that naturally secreted pheromone concentration is expectedly lower due to the inherent mechanism of physiological stimulation and thus, saturation phenomena are foreseen to be averted in that case. We also observed the Bar1p+ receiver exhibits its highest response within 6 minutes of induction, followed by a gradual decrease of transcriptional activity while \textit{bar1$\Delta$} cells maximize their transcriptional response after 30 minutes following initial induction. In contrast to the observations in Fig.\ref{Single Stimulation RNA}, over the physiological setting the two receivers exhibit distinct RNA upregulation profiles that imply that the \textit{bar1$\Delta$} deletion mutant can detect pheromone gradients more accurately when stimulated by non-saturating pheromone concentrations. This is further confirmed by the more robust response of these cells at timepoints 30' and 60', as compared to the less sensitive Bar1p+ strain. Collectively, our experimental data suggest that both yeast strains exhibit a more measurable and accurate response in the physiological stimulation setting which highlights their distinct profiles when it comes to reporting pheromone signalling.

\section{Conclusion}\label{conclusion}
In this paper, we present the first MC testbed using engineered yeast cells. The consideration of yeast cells is motivated by their genetic amenability, the wealth of knowledge on this model organism and its relevance to biomedical applications, bringing MC closer to a new generation of practical applications. We showcase this relevance, by demonstrating how this testbed can be used as baseline for developing and characterizing yeast biosensors and their integration to in-body sensor networks for health monitoring. We then present mathematical models that describe the system, which are validated to a good extent using the obtained experimental data. Although low event rates are reported when using common fluorescence detection techniques, obtained RNA levels indicate fast cell response to stimuli which can lay the foundation towards achieving faster rates. Such faster rates as well as practical yeast biosensing applications, will be sought in the near future. Beyond that, noise analysis coupled with an appropriate modulation and signal detection mechanism for communication optimization, are currently under investigation.

\appendices
\section{Receiver model}
\label{Appendix Rx}
\begin{flalign}
&\frac{d[\alpha]}{dt} = -v1 \label{alpha factor ODE}&\\
&\frac{d[Ste2]}{dt} = -v2 + v3 - v5 &\\ 
&\frac{d[Ste2active]}{dt} = v2-v3-v4& \\
&\frac{d[Sst2active]}{dt} = v37-v38 &\\
&\frac{d[G\alpha\beta\gamma]}{dt} = -v6+v9 &\\
&\frac{d[G\alpha_{GTP}]}{dt} = v6 - v7 - v8 &\\
&\frac{d[G\alpha_{GDP}]}{dt} = v7 + v8 - v9 &\\
&\frac{d[G\beta\gamma]}{dt} = v6 - v9 - v10 + v11+ v21 + v23 + v25 \nonumber\\ &+ v27 + v32 &\\
&\frac{dSte5}{dt} = - v12 + v13 + v17 + v21 + v23 + v25 + v27\nonumber\\& + v32 &\\
&\frac{dSte11}{dt} = - v12 + v13 + v17 + v21 + v23 + v25 + v27\nonumber\\& + v32 &\\
&\frac{dSte7}{dt} = - v14 + v15 + v17 + v21 + v23 + v25 + v27\nonumber\\& + v32 &\\
&\frac{d[Fus3]}{dt} = - v14 + v15 + v17 + v21 + v23 + v25 + v27 \nonumber\\ &- v29 + v30 + v33 &\\
&\frac{[dSte20]}{dt} = - v18 + v19 + v21 + v23 + v25 + v27 + v32 &\\
&\frac{d[A]}{dt} = v12 - v13 - v16 &\\
&\frac{d[B]}{dt} = v14 - v15 - v16 &\\
&\frac{d[C]}{dt} = -v10 + v11 + v16 - v17 &\\
&\frac{d[D]}{dt} = v10 - v11 - v18 + v19 &\\
&\frac{d[E]}{dt} = v18 - v19 - v20 - v21 &\\
&\frac{d[F]}{dt} = v20 - v22 - v23 &\\
&\frac{d[G]}{dt} = v22 - v24 - v25 &\\
&\frac{d[H]}{dt} = v24 - v26 - v27 &\\
&\frac{d[I]}{dt} = v26 - v28 + v31 &\\
&\frac{d[L]}{dt} = v28 - v29 + v30 - v32 &\\
&\frac{d[K]}{dt} = v29 - v30 - v31 &\\ 
&\frac{d[Fus3PP]}{dt} = v28 - v33&\\
&\frac{d[Bar1]}{dt} = -v36 + v37 &\\
&\frac{d[Bar1_{active}]}{dt} = v36 - v37 - v38 &\\
&\frac{d[Fus1_{mRNA}]}{dt} = P_3k_{trFUS1} - d_{mRNA}[Fus1_{mRNA}] \label{FUS1 gene}&\\
&\frac{d[Ste12]}{dt} = v39-v40 &\\
&\frac{d[Tec1]}{dt} = v41-v42 &\\
&\frac{d[SD1]}{dt} = v43 - v44 &\\
&\frac{d[SD2]}{dt} = v45-v46 &\\
&\frac{d[SD1D2]}{dt} = v47  &\\
&\frac{d[S2]}{dt} = v48-v49 &\\
&\frac{d[TS]}{dt} = v50-v51 &\\
&\frac{d[TSD1]}{dt} = v52-v53 &\\
&\frac{d[Ste12^*]}{dt} =v54-v55 &\\
&\frac{d[Tec1^*]}{dt} =v56-v57 &\\
&\frac{d[Fus1]}{dt} = v58-v59 \label{FUS1 ODE}&
\end{flalign}
Where,
\begingroup
\allowdisplaybreaks
\begin{flalign}
&v1 = [\alpha][Bar1_{active}] k_1 &\\
&v2 = [Ste2][\alpha]k_2 &\\
&v3 = [Ste2_{active}]k_3 &\\
&v4 = [Ste2_{active}]k_4 &\\
&v5 = [Ste2]k_5 &\\
&v6 = [Ste2_{active}][G\alpha \beta \gamma]k_6 &\\
&v7 = [G\alpha GTP]k_7 &\\
&v8 = [G\alpha GTP][Sst2_{active}]k_8 &\\
&v9 = [G\alpha GDP][G\beta\gamma]k_9 &\\
&v10 = [G\beta\gamma][C]k_{10} &\\
&v11 = [D]k_{11} &\\
&v12 = [Ste5][Ste11]k_{12} &\\
&v13 = [A]k_{13} &\\
&v14 = [Ste7][Fus3]k_{14} &\\
&v15 = [B]k_{15} &\\
&v16 = [A][B]k_{16} &\\
&v17 = [C]k_{17} &\\
&v18 = [D][Ste20]k_{18} &\\
&v19 = [E]k_{19} &\\
&v20 = [E]k_{20} &\\
&v21 = [E]k_{21} &\\
&v22 = [F]k_{22} &\\
&v23 = [F]k_{23} &\\
&v24 = [G]k_{24} &\\
&v25 = [G]k_{25} &\\
&v26 = [H]k_{26} &\\
&v27 = [H]k_{27} &\\
&v28 = [I]k_{28} &\\
&v29 = [L][Fus3]k_{29} &\\
&v30 = [K]k_{30} &\\
&v31 = [K]k_{31} &\\
&v32 = [L]k_{32} &\\
&v33 = [Fus3PP]k_{33} &\\
&v34 = [Ste12][Bar1]k_{36} &\\
&v35 = [Bar1_{active}]k_{37} &\\
&v36 = [Bar1_{active}]k_{38} &\\
&v37 = \frac{[Fus3PP]^2}{4^2+[Fus3PP]^2}k_{46} &\\
&v38 = [Sst2_{active}]k_{47} &\\
&v39 =  k_{s12}+k_{fb1}P_1 &\\
&v40 =  d_{s12}F_9[Ste12]-F_1-2F_2-F_3-F_5\nonumber &\\
&v41 =   k_{tec1}+k_{fb2}P_2 &\\
&v42 =  (d_{tec1}+J_1[Fus3PP]])[Tec1]- F_7\nonumber\\&-(F_5 + J_2[Fus3PP]])[TS] &\\
&v43 =  F_3+J_2[Fus3PP][TSD1] &\\
&v44 =  F_4+d_{sd1}F_9[SD1] &\\
&v45 = F_1 &\\
&v46 = F_7+d_{sd2}[SD2]&\\
&v47 = F_4 + F_7 &\\
&v48 = F_2 &\\
&v49 = d_{S2}F_9[S2] &\\
&v50 = F_5 &\\
&v51 = F_6+d_{ts}F_9[TS] &\\
&v52 = F_6 &\\
&v53 = d_{tsd1}F_9[TSD1] &\\
&v54 = 2d_{s2}F_9[S2]+
d_{sd2}F_9[SD2]+d_{tsd1}F_9[TSD1]\nonumber\\&+d_{s12}F_9[Ste12]+d_{ts}F_9[TS] &\\
&v55 = F_8[Ste12^*] &\\
&v56 = (d_{tec1}+J_1[Fus3PP])[Tec1]+\nonumber\\&(d_{ts}F_9+J_2[Fus3PP])[TS]+\nonumber\\&(d_{tsd1}F_9+J_2[Fus3PP])[TSD1] &\\
&v57 = F_8[Tec1^*] &\\
&v58 =  k_{trans}[Fus1_{mRNA}] &\\
&v59 = k_d[Fus1] &\\
&P_3 = \frac{[S2]}{[S2]+[Ste12]+[SD1]+KD3} &\\
&P_2 = \frac{[TS]}{[TS]+[TSD1]+KD2} &\\
&P_1 = \frac{[S2]}{[S2]+[Ste12]+[SD1]+KD1} &\\
&F_1 = k_c[Ste12](TDig2-uDig2)-\nonumber\\&(k_{\alpha}[Fus3PP] + kr_{sd2})[SD2] &\\
&F_2 = k_c[Ste{12}][Ste12]-d_s[S2] &\\
&F_3 = k_c[Ste12](TDig1-uDig1)-\nonumber\\&(k_\alpha[Fus3PP] + kr_{sd1})[SD1] &\\
&F_4 = k_c[SD1](TDig2-uDig2)-\nonumber\\&(k_\alpha[Fus3PP] + kr_{sd1d2})[SD1D2] &\\
&F_5 = k_c[Ste12][Tec1] - (J_2[Fus3PP] + kr_{ts})[TS] &\\
&F_6 = k_c[TS](TDig1-uDig1)-\nonumber\\&(k_\alpha[Fus3PP]+kr_{tsd1}) + J_2[Fus3PP])[TSD1] &\\
&F_7 = k_c[SD2](TDig1-uDig1)-\nonumber\\ &(k_\alpha[Fus3PP] + kr_{sd1d2})[SD1D2] &\\
&uDig1 = [SD1]+[SD1D2]+[TSD1] &\\
&uDig2 = [SD2]+[SD1D2] &\\
&F_8 = \frac{Km_{sat}}{[Ste12^*]+[Tec1^*]+KD_{sat}} &\\
&F_9 = \frac{k_{p1}[Fus3PP]}{[Fus3PP] + k_{p2}} + k_{p3} &
\end{flalign}
\endgroup\\

\setlength{\tabcolsep}{0.0001pt}
\topcaption{Chemical compounds involved in the Rx model}
\begin{supertabular}{|c||c|}

\hline
[$\alpha$]& $\alpha$ factor concentration [nM]\\
\hline
[Ste2]& receptor protein concentration [nM]\\
\hline
[$\mathrm{Sst2_{active}}$]& G protein regulator concentration [nM]\\
\hline
[$\mathrm{G_{\alpha \beta \gamma}}$]& G protein concentration [nM]\\
\hline
[$\mathrm{G_{\alpha}}$]& G protein subunit concentration[nM]\\
\hline
[$\mathrm{G_{\beta \gamma}}$]& G protein subunit concentration[nM]\\
\hline
[Ste5]& scaffold protein concentration[nM]\\
\hline
[Ste11]& scaffold protein concentration[nM]\\
\hline
[Ste7]& scaffold protein concentration[nM]\\
\hline
[Fus3]& MAPK concentration[nM]\\
\hline
[Bar1]& Bar1 concentration[nM]\\
\hline
[$\mathrm{Fus1_{mRNA}}$]& Fus1 mRNA concentration[nM]\\
\hline
[Ste12]& Ste12 transcription factor concentration[nM]\\
\hline
[Tec1]& Tec1 transcription factor concentration[nM]\\
\hline
[SD1]& Ste12-Dig1 complex concentration[nM]\\
\hline
[SD2]& Ste12-Dig2 complex concentration[nM]\\
\hline
[SD1D2]& Ste12-Dig1-Dig2 complex concentration[nM]\\
\hline
[S2]& Ste12 homodimer concentration[nM]\\
\hline
[TS]& Ste12-Tec1 heterodimer concentration[nM]\\
\hline
[TSD1]& Tec1-Ste12-Dig1 complex concentration[nM]\\
\hline
[Fus1]& Fus1 protein abundance [nM]\\
\hline
\end{supertabular}

\section*{Acknowledgment}
This research was supported by the projects EXCELLENCE/0421/0248 (PheroMolCom) and EXCELLENCE/0421/0302 (N-terDAM) which are implemented under the Cohesion Policy Funds "THALEIA 2021-2027" with EU co-funding.

\ifCLASSOPTIONcaptionsoff
  \newpage
\fi

\end{document}